\documentclass[prd,twocolumn,nofootinbib,groupedaddress,showkeys,citeautoscript]{revtex4}
\pdfoutput=1
\usepackage{graphicx}
\usepackage{epstopdf}
\usepackage{amsmath}
\usepackage{dcolumn}
\usepackage{multirow}
\usepackage{fancybox}
\usepackage{bm}
\usepackage{multirow}
\usepackage{array}
\usepackage{rotating}
\usepackage{ulem,fancyvrb}
\usepackage{latexsym}
\usepackage{amsfonts}
\usepackage{xcolor}
\usepackage{longtable}
\usepackage{amssymb}
\def\tana2{\tan^2\alpha_H}
\def\tanb2{\tan^2\beta}

\def\m0pr{m_0'}

\def\Mz2{M_Z^2}

\def\GeV{~{\rm GeV}}
\def\TeV{~{\rm TeV}}
\def\met100{\slashed{E}_T\geq 100~{\rm GeV}}

\newcommand{\beqn}{\begin{eqnarray}}
\newcommand{\eeqn}{\end{eqnarray}}
\newcommand{\be}{\begin{equation}}
\newcommand{\ee}{\end{equation}}
\def \cha{\tilde{\chi}^{\pm}_1}

\def \chpa{\tilde{\chi}^{+}_a}

\newcommand{\na}{\ensuremath{\tilde{\chi}^{0}_1}}
\newcommand{\noi}{\ensuremath{\tilde{\chi}^{0}_i}}
\def \nb{\tilde{\chi}^{0}_2}
\def \nc{\tilde{\chi}^{0}_3}

\def \n34{\tilde{\chi}^{0}_{3,4}}

\def \ta{\tilde{t}_1}
\def \tb{\tilde{t}_2}

\def \mhf{m_{1/2}}

\newcommand{\fb}{\ensuremath{\mathrm{fb}^{-1}}}
\newcommand{\sta}{\ensuremath{\tilde{\tau}_1}}

\DeclareMathOperator{\sgn}{sgn}

\allowdisplaybreaks[1]

\begin{document}

\title{ Naturalness,  Supersymmetry and Implications for LHC and Dark Matter}
\author{Sujeet~Akula}
\affiliation{Department of Physics, Northeastern University,
 Boston, MA 02115, USA}

\author{Mengxi~Liu}
\affiliation{Department of Physics, Northeastern University,
 Boston, MA 02115, USA}

\author{Pran~Nath}
\affiliation{Department of Physics, Northeastern University,
 Boston, MA 02115, USA}

\author{Gregory~Peim}
\address{Department of Physics, Northeastern University,
 Boston, MA 02115, USA}

 \begin{abstract}
It is shown that the Hyperbolic Branch  of the radiative electroweak symmetry breaking contains in it three regions: the Focal  Point, Focal Curves, and Focal Surfaces.  Further, the Focal Point  is shown to lie on the boundary of a Focal Curve. These focal regions allow for a small $\mu$ while scalar masses can become large and may lie in the several TeV region.  It is shown that for the mSUGRA model the current LHC-7 constraint depletes the Focal Point region  while regions on Focal Curves and Focal Surfaces remain largely intact. The LHC implications for models which lie on  Focal Curves are briefly discussed as well as the implications of dark matter constraints for the Focal~Point, Focal~Curves and Focal Surfaces are discussed. 
\end{abstract}


\maketitle



\section{Introduction \label{intro}}

Several naturalness, hierarchy, and, fine-tuning problems exist in particle physics: some big and some small.
The most severe one relates to the smallness of the vacuum energy in units of the Planck mass,  followed by 
the smallness of the ratio $M_W/M_{\rm Pl}$. There are several other small-to-moderate size
hierarchies such as the ratio $M_{\rm GUT}/M_{\rm Pl}$ and the ratios in the fermion mass spectra  such as
$m_u/m_t$.  Also, there are hierarchy problems of a  more technical nature, such as
 in the Higgs sector of  the standard model, where the Higgs boson
mass receives a loop correction which is quadratically dependent on the cutoff. 
 This problem is resolved in supersymmetric models with a cancellation 
 between the fermionic  and super-fermionic loops which results in the quadratic dependence on the cutoff being
 replaced by a logarithmic dependence.  
A similar problem at a much smaller scale often called the little hierarchy problem 
appears for supersymmetric
models  if the scalar masses turn out to be large. In fact, 
in certain models of soft breaking the  scalar 
masses   can get large, as is the case in 
supergravity grand unified models~\cite{Chamseddine:1982jx} with 
hierarchical breaking 
of supersymmetry~\cite{Kors:2004hz} and for certain string motivated models~\cite{Acharya:2011te}.  
Large scalar masses have also been considered in other contexts~\cite{Wells:2004di}.

The little hierarchy problem can be simply described as follows: in the radiative electroweak
symmetry breaking~(REWSB) one has $\frac{1}{2} M_Z^2\simeq- \mu^2 - m_{H_2}^2$ where $\mu$ is the Higgs mixing parameter
and $m_{H_2}$ is the mass of the Higgs boson that couples to the top quark.  Naively  $m_{H_2}$ gets large 
as the universal scalar mass  $m_0$ gets large 
and a large cancellation is needed between $\mu$ and $m_{H_2}$ to get a small
$M_Z$. A more practical approach is to view the REWSB relation as a determination of $\mu$ which is the 
view point we adopt here. From this perspective, if $m_0$ is large the accessibility of sparticles at the LHC 
rests on the size of $m_{1/2}$ and $\mu$ and thus a  small $\mu$ (and  a
small $m_{H_2}$) is desirable.
We note in passing that if $m_0$ is indeed large, the LHC would turn into a gaugino factory with the sparticles produced being gluinos~\cite{Feldman:2009zc}, charginos and neutralinos (see Sec.~\ref{s7}). We also note that this region gives a significant enhancement to proton lifetime~\cite{Review} because of the smallness
of the gaugino masses and relative heaviness of the squark masses.
 
The question then is how one may achieve a small $\mu$ for the above class of models 
in the context of radiative electroweak symmetry breaking. 
The basic mechanism for achieving the above was first realized in~\cite{Chan:1997bi}
(for further works see~\cite{HB2,bbbkt,Feldman:2008jy}).
In the analysis of~\cite{Chan:1997bi} it was found that there  exist two natural regions of
radiative breaking, one where there is an upper bound on the soft parameters $m_0$, $m_{1/2}$,  $A_0$ 
for a fixed $\mu$ (the Ellipsoidal Branch, EB), and the other where one or more soft parameters
can get very large  for fixed $\mu$ (the Hyperbolic Branch, HB).
 In a later work,~\cite{Feng:1999mn}, it was shown that there exists a  region where
the value of the Higgs mass squared, $m_{H_2}^2$, becomes essentially independent of the values
of the input parameter $m_0$ at the GUT scale. 
Such a region was then labeled the Focus~Point.

  In this work we classify the solutions of the Hyperbolic~Branch in Sec.~\ref{s2} and show that it contains 
three main regions:
(1)
 Focal~Points~(HB/FP):  This region lies at the  boundary between the Ellipsoidal and the   Hyperbolic Branches
 where $\mu^2$ becomes independent of $m_0^2$ and thus $m_0$ can get large while
 $\mu$ remains fixed with the other soft parameters being held fixed.  In this definition we do not include the Focal~Point on the EB. 
 The Focal~Point is  technically different from the Focus~Point~\cite{Feng:1999mn} but for $\tan\beta \gg 1$ they are essentially the same as will be made  clear in Sec.~\ref{s2} and Sec.~\ref{s3}. The 
 HB/FP region, however, is only a small part of HB  
 and the larger  parts of HB are Focal~Curves and Focal~Surfaces  as discussed below, and in detail in Sec.~\ref{s2} and Sec.~\ref{s4}.
 (2)
  Focal~Curves~(HB/FC): Focal~Curves  are where 
  two soft parameters are comparable and can get large while $\mu$ is fixed. We define HB/FC such that the HB/FP region is excluded.
   (3) 
  Focal~Surfaces~(HB/FS): Here one may have a fixed (and small) $\mu$ while the three dimensional  soft parameters may
  get large.  The HB/FS  region is the set of all Focal Curves and thus does not include the HB/FP region.
In Sec.~\ref{s7}, we carry out a numerical analysis of the mSUGRA parameter space under all the experimental constrains including the constraint from the recent LHC-7 data and analyze their effects on the HB region. 
We will show that the combined constraints severely deplete 
the Focal Point region, while the Focal Curves and thus Focal Surfaces largely remain intact. We 
also explore implications for SUSY discovery at the LHC and in dark matter searches.
Concluding remarks are presented in Sec.~\ref{s8}.


\section{Focal Points, Curves, Surfaces of the Hyperbolic Branch \label{s2}}
In this section we will discuss in detail the classification of HB into the three broad regions 
mentioned in the last section. 
We begin with the equation for the radiative breaking of the electroweak symmetry
\begin{equation} 
\mu^2 + \frac{1}{2} M_Z^2   =  \frac{\overline{m}_{H_1}^2 - \overline{m}_{H_2}^2 
\tan^2\beta}{\tan^2\beta - 1}~,
\label{1.1} 
\end{equation}
where  we have  $\overline{m}_{H_i}^2=m_{H_i}^2+\Sigma_i$ and 
  $\Sigma_i$ is the contribution arising from  the  loop corrections to the effective potential for $i=1,2$~\cite{Arnowitt:1992qp}.  In the analysis here we will focus on the supergravity grand unification model with universal boundary
  conditions~\cite{Chamseddine:1982jx,hlw,nac} whose soft breaking sector is described by 
  \beqn
\left(m_0,~m_{1/2},~A_0,~\tan\beta,~\sgn(\mu)\right)~,  
\label{universal}
  \eeqn
  where $m_0$ is the universal scalar mass, $m_{1/2}$ is the universal gaugino mass, $A_0$ is the
  universal trilinear coupling and $\mu$ is the Higgs mixing parameter in the
  superpotential.  The model of Eq.~\eqref{universal} is referred to as mSUGRA
   or sometimes as 
  the constrained minimal supersymmetric model, CMSSM. 
  The analysis is done using the techniques given in~\cite{an1992} where one starts with universal boundary conditions given by Eq.~\eqref{universal}
   for the soft parameters at the GUT scale and evolves the sparticle masses downwards  using renormalization group 
 equations. 
 For illustration in the text, we consider one loop evolution where we neglect the Yukawa couplings except for the top 
 quark.
The simulations presented later are done using numerical codes which include the 
 effects of the $b$ and $\tau$ Yukawa couplings.
  As discussed in Sec.~\ref{intro}, the radiative electroweak symmetry breaking allows for a determination of  $\mu^2$ in terms of the 
  soft parameters as~\cite{Nath:1997qm,Chan:1997bi}
 \beqn
 \mu^2  = &-&\frac{1}{2}M_Z^2 +  m^2_0  C_1+ A^2_0 C_2\nonumber\\
 & +&
 m^2_{1/2} C_3+ m_{1/2}
A_0 C_4+ \Delta \mu^2_{\rm loop}~,
\label{1.6}
\eeqn
where 
\beqn \label{1.61}
C_1&=&\frac{1}{\tan^2\beta-1}\left(1-\frac{3 D_0-1}{2}\tan^2\beta\right)~,\\
\label{1.61a}
 C_2&=&\frac{\tan^2\beta}{\tan^2\beta-1}k~,\\
 \label{1.61b}
C_3&=&\frac{1}{\tan^2\beta-1}\left(g- e\tan^2\beta \right)~, \\
C_4&=&-\frac{\tan^2\beta}{\tan^2\beta-1}f~,
\label{1.8}
\eeqn
 and $e.f,g,k$ are as defined in~\cite{Ibanez:1984vq}. $D_0(t)$ is  defined by 
\beqn
 D_0(t)= \left(1+ 6 Y_0 F(t)\right)^{-1}~.
 \label{d0EQ}
 \eeqn
Here  $Y_0 =h_t(0)^2/(4\pi^2)$, where $h_t(0)$ is the top Yukawa coupling at the GUT scale, $M_G\simeq2\times10^{16}\GeV$. Further, 
$F(t) = \int_0^t E(t^\prime) dt^\prime~,$ where  
$E(t)=\left(1 + \beta_3 t\right)^{16/3b_3} \left(1+ \beta_2 t\right)^{3/b_2} \left(1+ \beta_1t\right)^{13/9 b_1}$.
Here 
$\beta_i = \alpha_i(0) b_i/(4\pi)$ and $b_i=(-3, 1, 11)$ for $SU(3), SU(2)$ and $U(1)$
 and $t= \ln \left(M_G^2/Q^2\right)$ where $Q$ is the
 renormalization group point.  Our normalizations are such that $\alpha_3(0) = \alpha_2(0) = \frac{5}{3} \alpha_1(0)
 =\alpha_G(0)$. 
 Further, $\Delta \mu^2_{\rm loop}$ is the loop correction~\cite{Arnowitt:1992qp}.
  
As is well known, the tree value of $\mu^2$ (Eq.~\eqref{1.6} without $\Delta \mu^2_{\rm loop}$), is sensitive to the renormalization group
scale and the same is true of the loop correction. However, the  sum of the tree and the loop term
is  relatively insensitive to variations in $Q$~\cite{Chan:1997bi}.
 Further, one of the interesting phenomenon observed in~\cite{Chan:1997bi} is the following: suppose one goes to a renormalization group point $Q$ where 
the loop contribution $\Delta \mu^2_{\rm loop}$ is minimized
(this typically occurs at $Q\sim\mathcal{O}\left(\sqrt{M_{\ta}M_{\tb}}\right)$, where $M_{\ta}$ and $M_{\tb}$ are the stop masses).  
Now at low values of $\tan\beta$ and $Q$ it is observed that the co-efficients $C_i$ ($i=1-4$) 
continue to be all positive. In this case it is clear that for any fixed $\mu$ the soft parameters have
 well defined upper limits. However,  for larger values of $\tan\beta$, $C_1$ can vanish or even 
turn negative as $Q$ increases.   We will call the region where $C_1$ either vanishes or is negative
as the Hyperbolic~Branch.  
In this case it is possible to have large soft parameters while $\mu$ remains relatively small.

\begin{figure}[t!]
\begin{center}
{ \bf \scriptsize  Domain of EB and of HB\\
 HB $\supset$ Focal~Point, Focal Curves,  
Focal Surfaces
}\\
\includegraphics[scale=0.13]{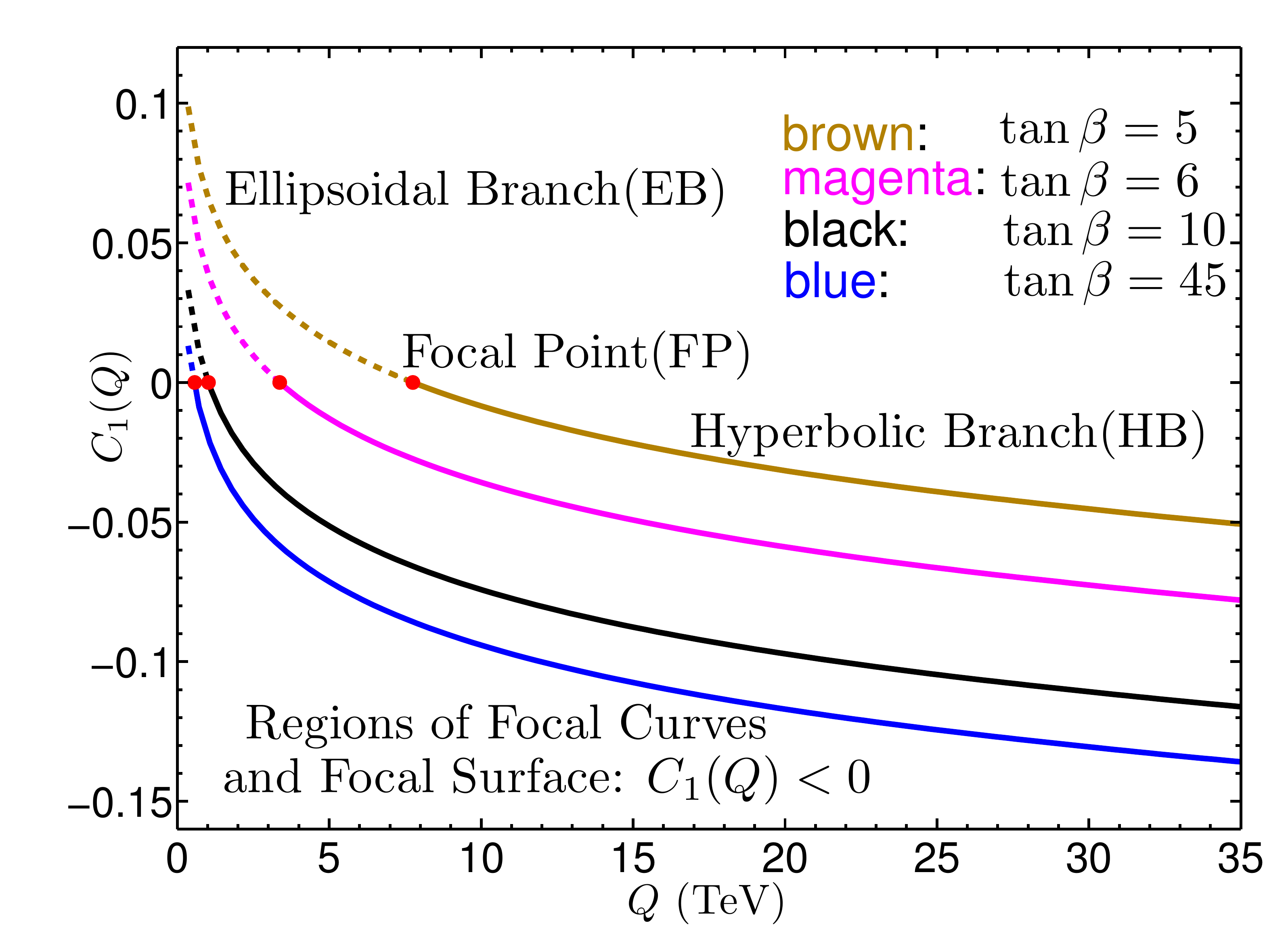}
\includegraphics[scale=0.13]{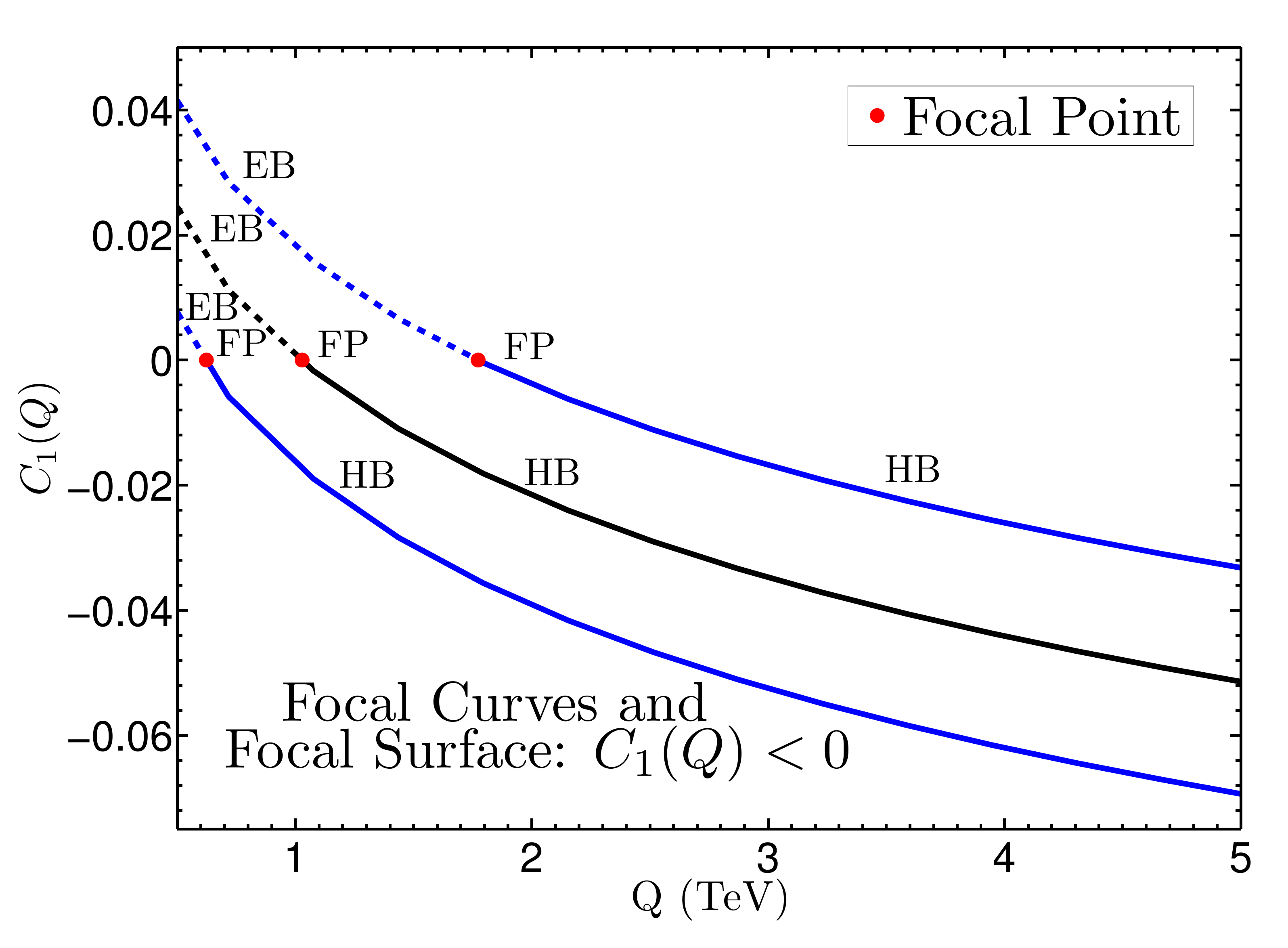}
\caption{\label{c1_c2_fig} 
Left:
A display of $C_1$ as a function of $Q$ 
 for different values of $\tan\beta$, i.e.,  
 $\tan\beta=5$~(brown), $\tan\beta=6$~(magenta), $\tan\beta=10$~(black) and $\tan\beta=45$~(blue). 
For larger values of $\tan\beta$, $C_1$ is positive for $Q\lesssim 1\TeV$.
Right: A display of the sensitivity of $C_1(Q)$ to the top quark  mass.  
The blue lines correspond to $\pm  1 \sigma$ in the top pole mass 
around the black line which corresponds to the central value, 
where the pole mass is  taken to be $m_t=\left(173.1\pm1.3\right)\GeV$. 
  \label{focalpoint}
}
\end{center}
\end{figure}

The HB of REWSB contains three regions:
(1) The Focal~Point (HB/FP): 
 We define the points where $C_1$ vanishes as Focal Points. 
 From Eq.~\eqref{1.6} and Eq.~\eqref{1.61} we find  that when $C_1=0$, 
  $m_0$ can get large without affecting $\mu$.  For practical purposes, for a fixed $\tan\beta$, we will 
    take a small region around $C_1=0$, and call it the Focal Point region, specifically
    \beqn
|C_1| <   \delta\left(Q,m_{t}\right), ~~~\delta\left(Q,m_{t}\right)\ll1~.
\label{2.9}
\label{lhc2}
\eeqn
In determining  $\delta\left(Q,m_{t}\right)$ we are guided by the experimental error in the top quark mass from 
$m_t = (173.1\pm 1.3)\GeV$.  Now, for a fixed $\tan\beta$, $C_1=C_1(m_t,Q)$ where $Q\sim\mathcal{O}\left(\sqrt{M_{\ta}M_{\tb}}\right)$ 
and thus, $Q$ depends on the top mass via the dependence of the stop masses on $m_t$. However, this
implicit dependence on $m_t$ via $Q$ is rather weak and effectively 
$\delta C_1 = \frac{\delta C_1}{\delta m_t}\delta m_t$.
A direct analysis gives the following approximate result
\beqn
\delta C_1 \simeq 3\left(1-D_0\right) \frac{\delta m_t}{m_t}~.
\label{deltac1}
\eeqn
This result agrees with the one loop  analysis in Fig.~\ref{c1_c2_fig} where $\delta C_1$
 can be interpreted as the vertical spacing between the curves in the right panel of Fig.~\ref{c1_c2_fig}.
In the full numerical analysis presented later in identifying the parameter points that lie in the Focal
Point region, we calculate $\delta C_1$ numerically for each point by calculating the variation in $C_1$
for variations in $m_t$.
  (2) Focal~Curves (HB/FC): The region where $C_1<0$ allows for two soft parameters  to get large while $\mu$ remains small is
the Focal Curve region. In fact, in this case there are two general possibilities: HB/FC1 and HB/FC2.
 In the case of HB/FC1~(HB/FC2), we have $C_1<0$ and $m_{1/2}$~($A_0$) as well as $\mu$ are held fixed with $m_0$ and
$A_0$~($\mhf$) allowed to vary.
These two cases can be combined into a single form HB/FC$_{\alpha}$ defined by $C_1<0$ and
the constraint $\left(1-\alpha\right)m_{1/2} = \alpha \left|A_0\right|$ where $0<\alpha <1$.
We note that HB/FC$_{\alpha}$ reduces to HB/FC1 when $\alpha\sim 0$ and 
reduces to HB/FC2 when $\alpha \sim 1$. 
 (3)  The Focal~Surface  (HB/FS): is the region of HB where 
$C_1< 0$ while all the soft parameters (except $\tan\beta$), i.e, $m_0, m_{1/2}, A_0$ vary and may 
 get large while $\mu$ remains fixed. In terms of $\mathrm{HB/FC}_\alpha$, varying $\alpha$ creates a Focal Surface.
 
We discuss now briefly the issue of fine-tuning. 
Often one uses the criterion of fine-tuning to designate 
some regions of the  parameters as preferred over others. 
However, such criteria are necessarily subjective and widely different results  can be
attained by different choices. 
For example, one criteria
used is to look at the sensitivity of $M_Z$ to variations in the  parameters that enter in 
Eq.~\eqref{1.1}.  Let us define  the set of such parameters to be $a_i$, then the sensitivities 
$f_i$ and the fine tuning parameter $f$ 
are taken to be as in \cite{Feng:1999mn}:
\beqn
f_i = \left |\frac{\partial \ln M_Z^2}{\partial \ln a_i}\right |~, ~~f=\max \{f_i\}~.\label{s1}
\eeqn
(For some early works related to naturalness and fine-tuning 
see~\cite{bg,ACR,Kane:1998im,Chankowski:1998xv,Casas:2003jx} and for more recent works see \cite{Martin:2010kk,Feldman:2011ud,afs,amsy}.)

Using the above criteria it has been argued that certain regions of the parameter space (such as
when $A_0\neq 0$) are less natural than the $A_0=0$ region~\cite{Feng:1999mn}.
However, such an argument appears to us as hasty in suppressing parts of the parameter
space based purely on a theoretical prejudice. For example, as already noted in~\cite{Feng:1999mn}
inclusion of the top Yukawa in the list  $\{a_i\}$ would lead to very different conclusions.
It may turn out that nature chooses a parameter point which one might consider `unnatural' from
a criteria such as of Eq.~\eqref{s1} but is perhaps the natural consequence of a more unified 
approach.
In our analysis we will not rely on criteria such as Eq.~\eqref{s1} as a  selection principle for 
the parameter space.
Such criteria are subjective and thus weaker.
Rather, we take the more pragmatic
approach regarding exploration of the entire parameter from a phenomenologically desirable view point.
Such a  view point requires that we  explore the small $\mu$ region of the  parameter space
while one or more of the other soft parameters  (such as $m_0$ and $A_0$) could become large.
Effectively our naturalness criteria will be simply regions of small $\mu$ as in the
analysis of~\cite{Chan:1997bi}.
Thus solutions of this type appear desirable  for phenomenological
reasons regarding the detectability of new physics at the LHC.  Further, as mentioned earlier situations of this type arise in theory models~\cite{Kors:2004hz,Acharya:2011te}.



 \section{The Focus Point region of  HB \label{s3}}
 While the Hyperbolic Branch~\cite{Chan:1997bi} and the Focus Point~\cite{Feng:1999mn} 
 both allow for large values of $m_0$ while
 $\mu$ remains small, the exact relationship of the Hyperbolic Branch and of the Focus Point
 has not been elucidated in the literature.  In this section we establish a direct connection between the two.
  We show that the Focus Point is
 the boundary point of a Focal Curve on the Hyperbolic Branch. 
Again for illustration we will consider one loop evolution, and among the Yukawa couplings
retain only the top quark coupling. 
 Here the scalar masses 
$m_{H_2}^2$,  $m_{\tilde U}^2$ and  $m_{\tilde Q}^2$ satisfy 
 the following set of coupled equations 
\begin{eqnarray}
\frac{dm_{H_2}^2}{dt}&=&-3Y_t\Sigma  - 3Y_t A_t^2
+\left(3\tilde\alpha_2M_2^2+
\tilde\alpha_1M_1^2\right)\label{dmh2}~,\nonumber\\
\frac{dm_{\tilde U}^2}{dt}&=&-2Y_t\Sigma -2 Y_t A_t^2
+\left(\frac{16}{3}\tilde\alpha_3M_3^2
+ \frac{16}{9} \tilde\alpha_1M_1^2\right)~\label{dmu2},\nonumber\\
\frac{dm_{\tilde Q}^2}{dt}&=&-Y_t\Sigma -Y_t A_t\nonumber\\
&~+&\left(\frac{16}{3}\tilde\alpha_3M_3^2
+ 3\tilde\alpha_2M_2^2+
\frac{1}{9} 
\tilde\alpha_1M_1^2\right)~\label{dmq2},~
\end{eqnarray}
where $\Sigma=(m_{H_2}^2+m_{\tilde Q}^2+m_{\tilde U}^2)$,
$Y_t= h_t^2/(16\pi^2)$, and where $h_t$ is the Yukawa coupling at scale $Q$. 
The analysis of~\cite{Feng:1999mn} made the  observation that the 
solution to Eq.~\eqref{dmh2}, 
 can be written in the form $m_i^2= (m_i^2)_p + \delta m_i^2$ where 
$(m_i^2)_p$ is the particular solution and the $\delta m_i^2$ obey 
the  homogeneous equation
\beqn
\frac{d}{dt}\left[\begin{array}{ccc}
        \delta m^2_{H_2}           \\
        \delta m_U^2    \\
        \delta m_Q^2        
	\end{array}\right] 
= - Y_t  \left[\begin{array}{ccc}
        3      &  3 & 3           \\
        2      &   2 &   2  \\
        1 &  1 & 1 \\
	\end{array}\right]
\left[\begin{array}{ccc}
       \delta m^2_{H_2}           \\
        \delta m_U^2    \\
        \delta m_Q^2        
	\end{array}\right]~. 
  	\eeqn
The solution to the above with the universal boundary conditions at the GUT scale is given by 
\beqn
\left[\begin{array}{c}
        \delta m^2_{H_2}           \\
        \delta m_U^2    \\
        \delta m_Q^2        
	\end{array}\right] 
= \frac{m_0^2}{2}   \left[\begin{array}{c}
        3J(t)-1          \\
        2 J(t)~~~~~ \\
        J (t) +1 
	\end{array}\right]~,
	\eeqn
where $J$ is an integration factor defined by 
\beqn
J(t) \equiv  \exp\left[-6\displaystyle\int_0^{t} Y_t(t') dt'\right]~.
\label{j1}
\eeqn
As $Q \to M_G$, one has $J(t)\to 1$ and the universality of the masses is recovered at the GUT scale.
Noting that  $Y(t)$ at the one loop level satisfies the equation
\beqn
\frac{dY_t}{dt} = \left(\frac{16}{3} \tilde \alpha_3 + 3 \tilde\alpha_3+\frac{13}{9} \tilde \alpha_1\right) Y_t - 6 Y_t^2~,
\eeqn
one finds $Y_t$ so that
\beqn
Y_t(t) = \frac{Y(0) E(t)}{1 + 6 Y(0) F(t)}~.
\label{aj2}
\eeqn
where $F(t)$ and $E(t)$ are defined after Eq.~\eqref{d0EQ}, one can inspect $J(t)$ to find that
$J(t) = D_0(t)$,
where $D_0(t)$ is defined by Eq.~\eqref{d0EQ}. 
Thus   $\delta m^2_{H_2}$ takes the form 
   \beqn
\delta\overline{m}_{H_2}^2 \equiv \frac{\delta m^2_{H_2}}{m_0^2}   = \frac{1}{2} \left(3D_0-1\right)~.
  \label{j3}
\eeqn
and $C_1$ can be expressed in terms of $\delta\overline{ m}_{H_2}^2$
\beqn
C_1= \frac{1}{\tan^2\beta-1}\left(1- \delta\overline{ m}_{H_2}^2\tan^2\beta\right)~.
\label{j4}
\eeqn
From Eq.~\eqref{j3}  we see that the correction  $\delta m^2_{H_2}$  becomes independent of 
$m_0$ when $D_0=1/3$, which corresponds to the so called Focus~Point~region~\cite{Feng:1999mn}, and from Eq.~\eqref{j4} one finds that $\delta\overline{ m}_{H_2}^2 \to 0$ implies that $C_1$ also vanishes, for $\tan\beta \gg 1$.  
Thus for large $\tan\beta$, i.e. $\tan\beta\gtrsim 5$, the Focal~Point and the Focus~Point essentially merge.  More explicitly, the Focus~Point implies the vanishing of $\delta\overline{ m}_{H_2}^2$ while the Focal~Point requires the vanishing of $C_1$.
 A numerical analysis of the behavior of $C_1$ 
 as a function of
 $Q$ for a set of fixed $\tan\beta$'s is given in Fig.~\ref{focalpoint} as well as a graphical representation of the different branches. 
 Fig.~\ref{focalpoint} 
  shows that the Focal Point is the boundary point of HB or, in other words, the transition point between EB and HB. 
  

\section{ Focal~Curves  and Surfaces
 \label{s4}}
\begin{figure}[t!]
\begin{center}
{\scriptsize \bf  Evolution of $\sqrt{{|C_1|}/{C_2}}$  and $\sqrt{{|C_1|}/{C_3}}$ with $Q$}\\
\includegraphics[scale=0.18]{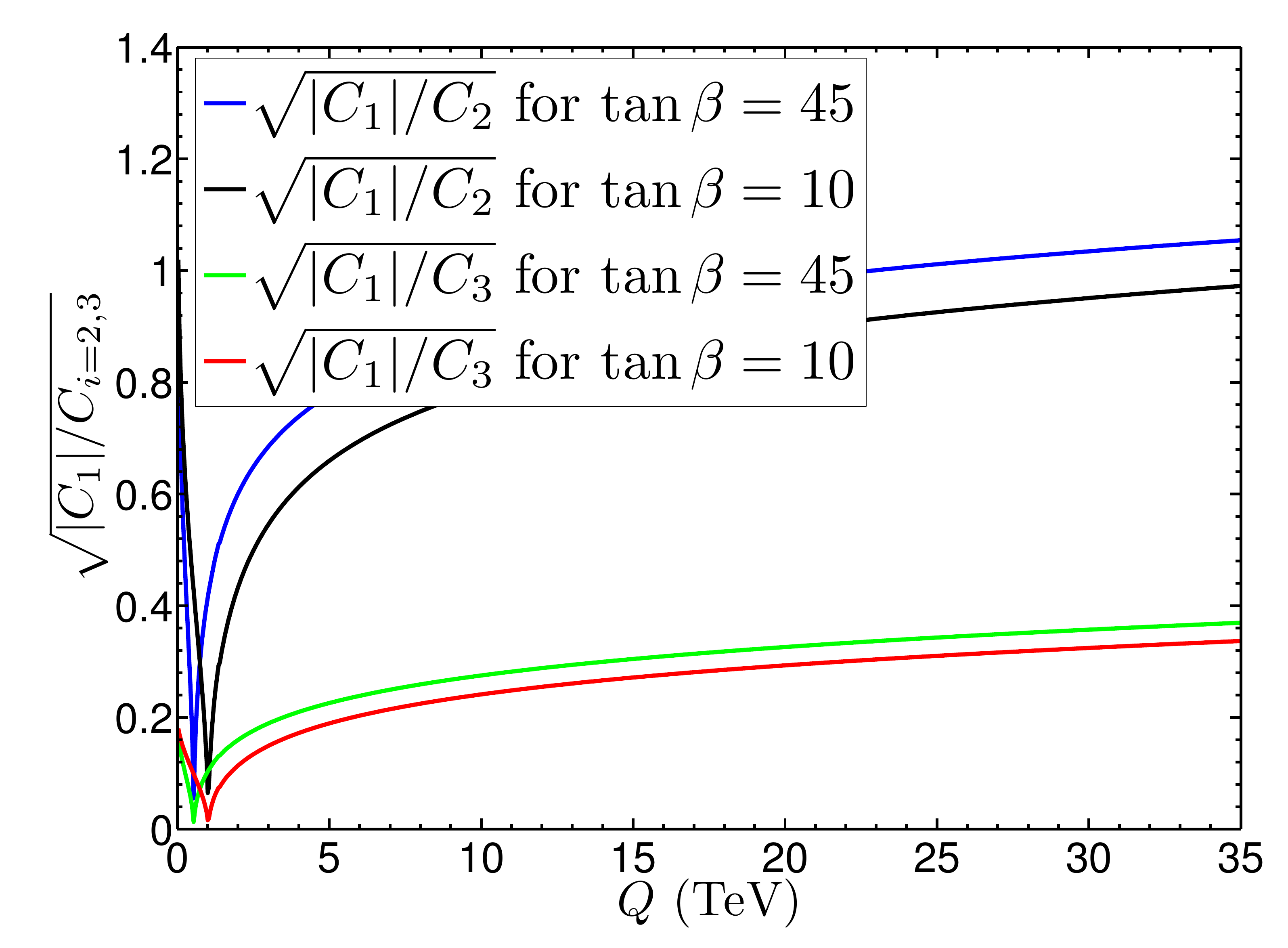}
\caption{
A numerical analysis of the evolution of 
$\sqrt{{|C_1|}/{C_2}}$  and $\sqrt{{|C_1|}/{C_3}}$ 
using Eq.~\eqref{1.61}, Eq.~\eqref{1.61a}, and Eq.~\eqref{1.61b}. Here one finds that 
$\sqrt{{|C_1|}/{C_2}}$  tends to $\sim1$ 
 and $\sqrt{{|C_1|}/{C_3}}$ tends to $\sim0.4$ as $Q$ becomes large.
 The analysis is shown for $\tan\beta=10$ and $\tan\beta=45$. 
\label{figc12c13} }  
\end{center}
\end{figure}

{\it Focal ~Curves (HB/FC):}
 To exhibit the emergence of a Focal Curve we rewrite Eq.~\eqref{1.1} in the following form
\beqn
 \mu^2  =- \frac{1}{2} M_Z^2 +  m_0^2 C_1 +  \overline{A}_0^ 2 C_2 + &m_{1/2}^{2} \overline{C}_3  + \Delta \mu^2_{\rm loop}~,
  \label{4.1a}~~\\
\overline{A}_0 \equiv  A_0+   \frac{C_4}{2C_2}  \mhf, ~&\overline{C}_3\equiv C_3- \frac{C_4^2}{4C_2}~.
 \label{4.2}
 \eeqn
  Now, suppose we go to the renormalization group point $Q$ where the loop corrections are 
  minimized and, further,  we are in a region of $\tan\beta$ and $Q$ where 
   $C_1$ is  negative. In this case one finds that 
     there exist curves where $m_0$ and $A_0$ get large while $m_{1/2}$ is held fixed
  and $\mu$ is relatively small compared to $m_0$ and $A_0$. 
   Thus  we can rewrite Eq.~\eqref{4.1a} in the form   
   \beqn
   \left(\overline{A}_0\sqrt{C_2} \right)^2 - \left(\sqrt{|C_1|} m_0\right)^2  = \pm|\mu_1|^2~~~{\rm HB/FC1}~,   
   \label{4.3} 
   \eeqn
  where 
$\pm |\mu_1|^2 \equiv \mu^2 + \frac{1}{2} M_Z^2 -  m_{1/2}^2 \overline{C}_3 - \Delta \mu^2_{\rm loop}$~,
   where $\pm$ indicates the overall sign of the right hand side. Thus one has two 
   branches corresponding to the two signs.
    We  can interpret Eq.~\eqref{4.3} 
   as an equation of a  Focal~Curve in the $m_0-\bar A_0$~plane (or in the $m_0-A_0$~plane around a shifted origin in $A_0$) such that as $m_0$ and $A_0$ get
large, $\mu$ remains fixed for fixed $m_{1/2}$~(this is Focal~Curve HB/FC1 as 
defined in Sec.~\ref{s2}). 
 In the limit when $m_0, |A_0|$ (and $Q$)  are much
larger than $\mu$ and $m_{1/2}$ 
one gets the result 
  \beqn
  \frac{\bar A_0}{m_0} \longrightarrow ~ \frac{A_0}{m_0} \longrightarrow \pm \sqrt{\frac{\left|C_1\right|}{C_2}} \longrightarrow \sim \pm  1~.
  \label{4.5aa}
  \eeqn
    where the last entry  in Eq.~\eqref{4.5aa} arises from a numerical evaluation of $C_1$ and $C_2$ as given 
  by Eq.~\eqref{1.61} and Eq.~\eqref{1.61a} as shown in Fig.~\ref{figc12c13}.

\begin{figure}[t!]
\begin{center}
\begin{tabular}{cc}
{\scriptsize \bf Focal~Curve HB/FC1} & {\scriptsize \bf Focal~Curve HB/FC2 }\\
\includegraphics[scale=0.13]{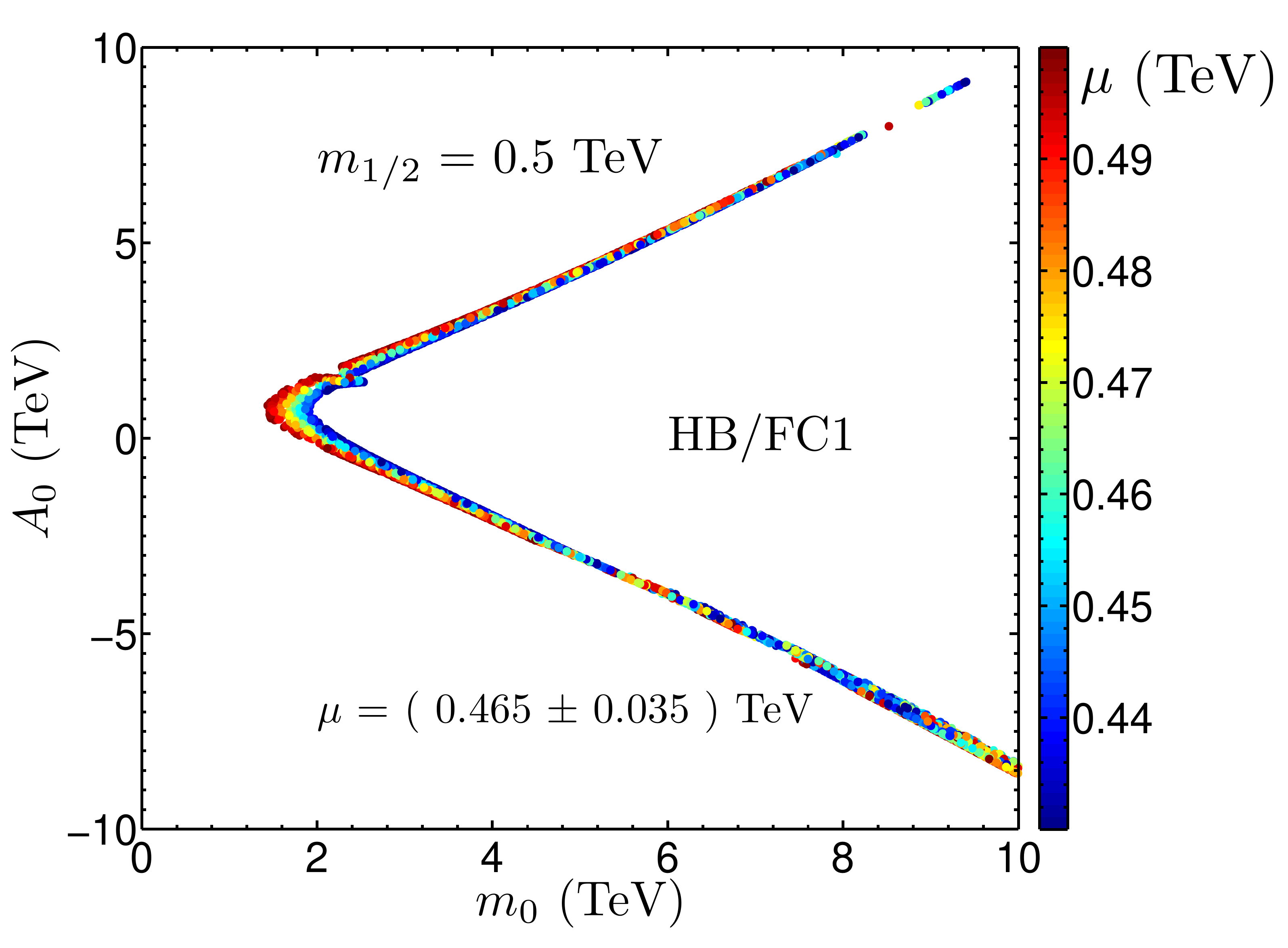}
&\includegraphics[scale=0.13]{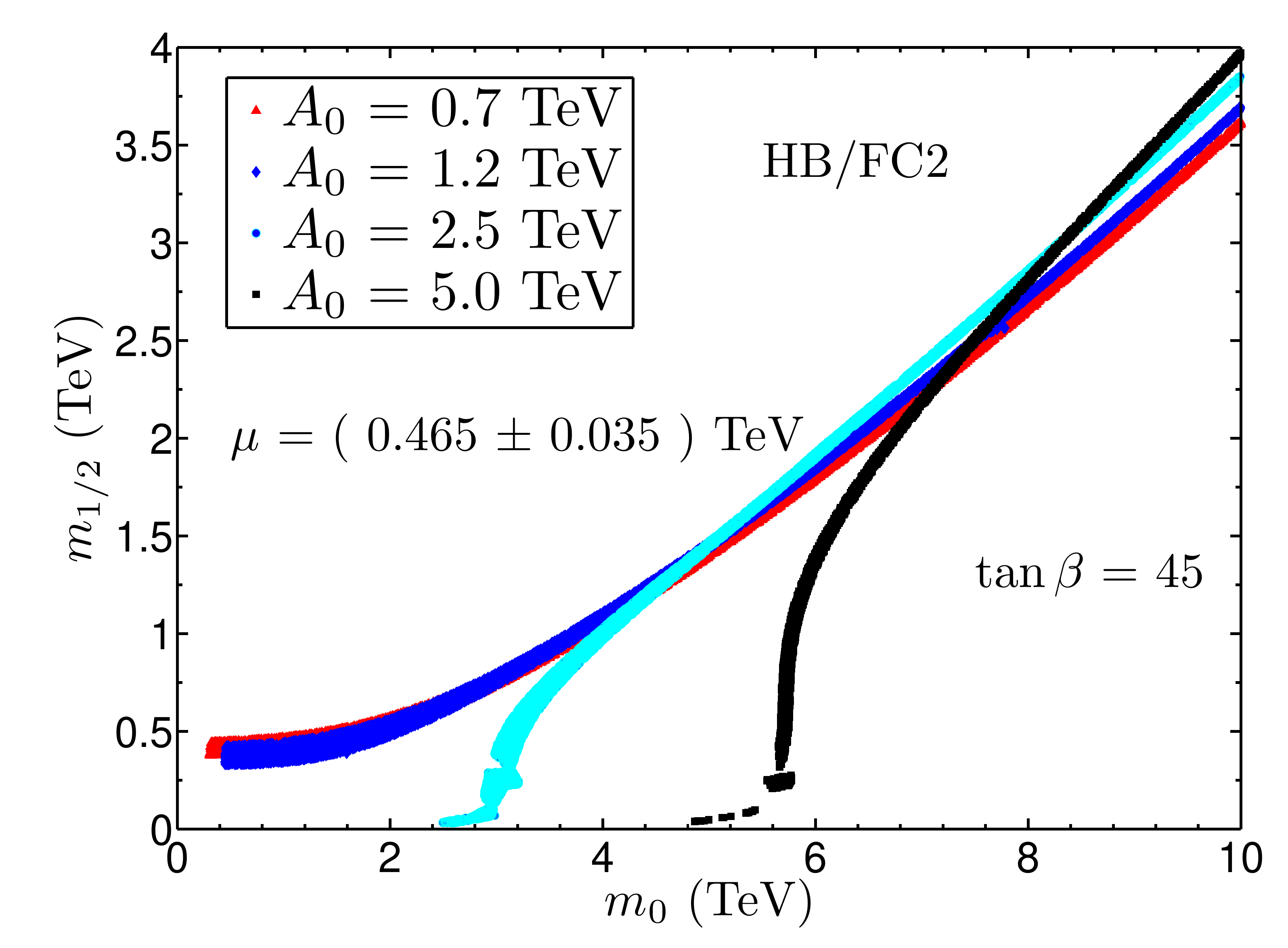}
\end{tabular}
\caption{
Left: Exhibition of  Focal~Curves HB/FC1 with $\mhf=0.5\TeV$ and $\tan\beta=45$ where $\mu$ lies in the range 
$\mu=(0.465\pm 0.035)\TeV$.  Points are displayed by $\mu$ value. 
Right: 
An illustration of Focal~Curves HB/FC2  which arise when $m_0$ and  $\mhf$ 
 are free to vary while $A_0$ is fixed and $\mu$ is held relatively constant. The analysis is  for $\tan\beta=45$ and for four values of
 $A_0$ which are  $A_0=0.7\TeV$~(red), $A_0=1.2\TeV$~(blue), $A_0=5.0\TeV$~(cyan) and $A_0=2.5\TeV$~(black). 
 The analysis above shows that on the Focal~Curve HB/FC1 and HB/FC2 one has good agreement with the asymptotic behavior as predicted by  Eq.~\eqref{4.5aa} and Eq.~\eqref{4.5a}.
 \label{a0_m0_nl}
 }
\end{center}
\end{figure}

\begin{figure}[t!]
\begin{center}
{ \bf \scriptsize  $m_0$ Reach for small $\mu$  on Focal Curves}\\
\includegraphics[scale=0.13]{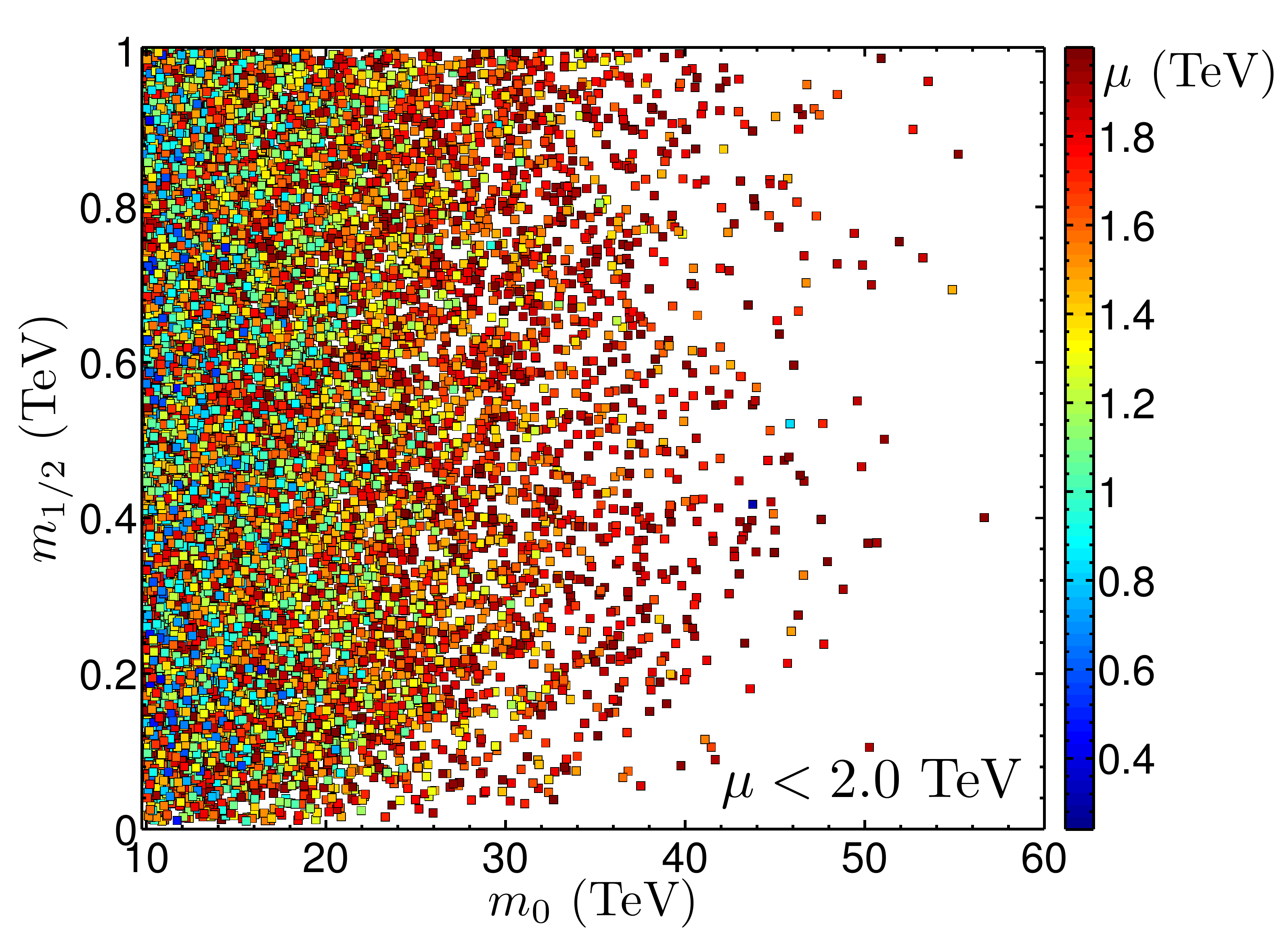} 
\includegraphics[scale=0.13]{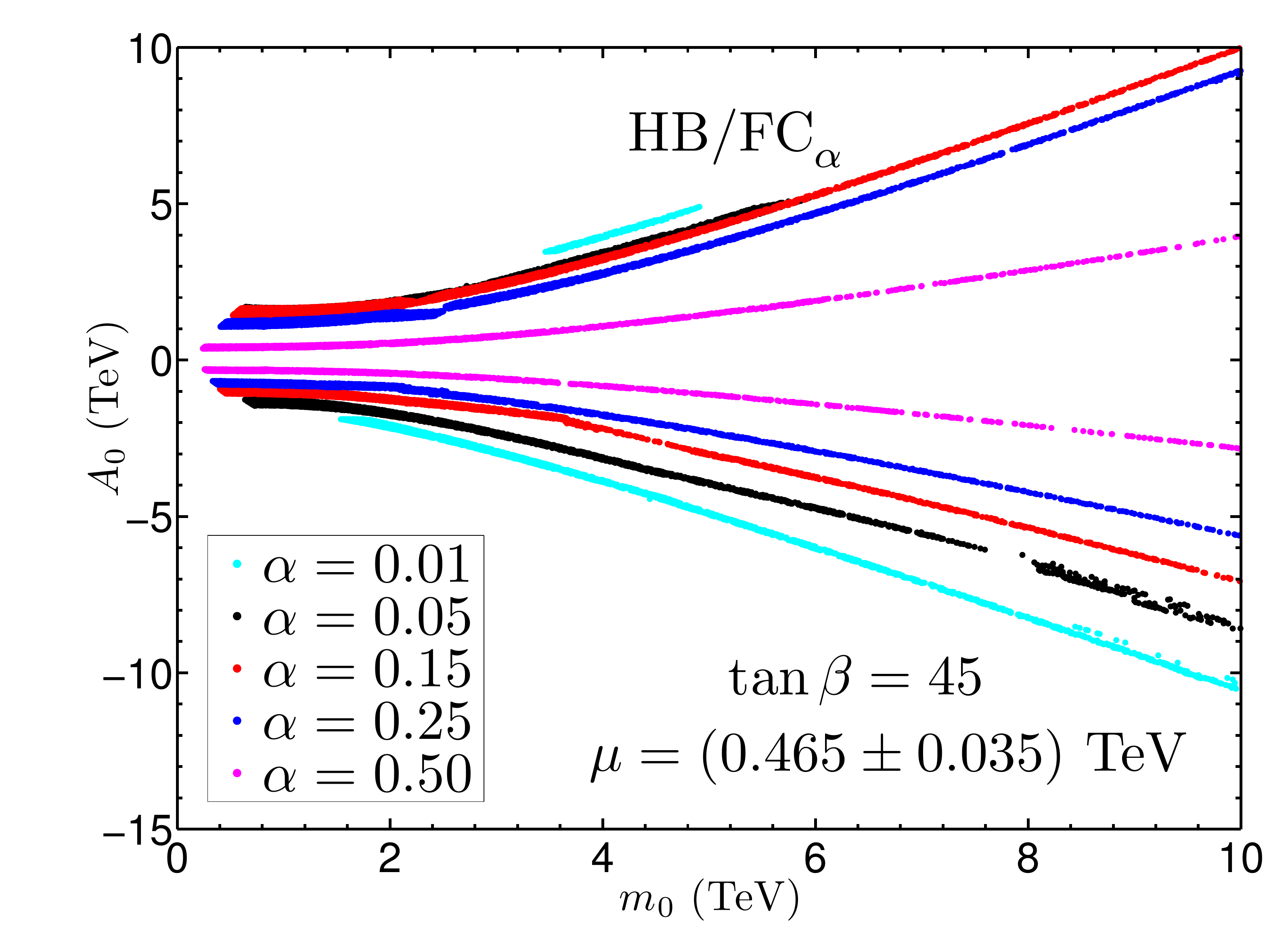}
\caption{Left:\label{largem0}
An exhibition of the reach in $m_0$ on Focal Curves HB/FC1 and HB/FP with $\mu<2\TeV$ consistent with radiative breaking of the
electroweak symmetry 
where points are displayed by their $\mu$ value in units of $\TeV$. It is seen that an $m_0$ as large as 
$50\TeV$  and above can be reached in this region.  Essentially all models lie on HB/FC1, but there are a few ($0.1\%$ of the displayed models) that are HB/FP.  Models were found by doing a uniformly distributed parameter scan of $m_0$, $\mhf$, $A_0$ and $\tan\beta$.
Right:\label{alph_fig} Exhibition of Focal~Curves HB/FC$_{\alpha}$ using 
$\mhf=\frac{\alpha}{1-\alpha}|A_0|$ for $\tan\beta=45$ and $\mu=\left(0.465\pm0.035\right)\TeV$ with $m_0$ between $10\GeV$ and $10\TeV$ and $A_0$ between $-8m_0$ and $8m_0$.  We display the cases where $\alpha=0.01,0.05,0.15,0.25,0.50$ and notice that for smaller $\alpha$ the asymptotic behavior is more steep. 
} 
\end{center}
\end{figure}

   In order to identify which points lie on  Focal~Curves we compute the 
   $C_i$ for each point and then subject them to the conditions necessary for
   them to lie on a Focal Curve. Thus for the case presented above we consider
   $m_{1/2}$  fixed while $m_0$ and $A_0$ vary with $C_1<0$ and outside the
   Focal Point region.  
     An analysis illustrating Focal~Curves in this case is given in Table~\ref{tab_a0_m0}. 
     For this analysis and subsequent figures and tables we use both {\sc SuSpect}~\cite{SuSpect} and {\sc SoftSUSY}~\cite{softsusy} which include 
      the two loop renormalization group equations and the two loop corrections to the Higgs sector.
    The analysis is done for the case when 
  $m_0$ lies in the range $500\GeV$ to $4000\GeV$ and $A_0$ lies in the range  $-500\GeV$ to $-3000\GeV$ with $\tan\beta=15$ and $\mu$ remaining within $10\%$ of  $600\GeV$.    
  A similar analysis is shown pictorially in the left panel of Fig.~\ref{a0_m0_nl}, where we have displayed the Focal~Curves for $\mhf=500\GeV$, $\tan\beta=45$ and $\mu=\left(465\pm35\right)\GeV$.  
  We see that for $m_0$ and $|A_0|$ large, there is good agreement with Eq.~\eqref{4.5aa}, i.e., one finds
  $A_0/m_0\to \pm 1$ asymptotically for large $m_0$. 
We note that the limit $A_0/m_0\sim 1$ consistent with small $\mu$ was noticed and discussed 
recently  in the analysis of~\cite{Feldman:2011ud} in the context of a string motivated model. From the left panel of Fig.~\ref{a0_m0_nl} we note that this limit is part of HB and
 is specifically the end point of the Focal~Curve HB/FC1.
   The left panel of Fig.~\ref{largem0} shows model points with $\mhf<1\TeV$ and $m_0>10\TeV$ with $\mu<2\TeV$.  The result of $m_0$ up to $10\TeV$ were exhibited in~\cite{Chan:1997bi}, and up to $30\TeV$ in~\cite{Feldman:2011ud}, 
   and 
 here we exhibit $m_0$ up to $50\TeV$ and beyond for $\mu<2\TeV$, i.e., $\mu / m_0 \ll 1$.
\begin{table}
 \begin{center}
 \begin{tabular}{|c|c|c|c|}
 \hline 
$m_0$~(GeV) & $A_0$~(GeV) & Q~(GeV) & $\mu$~(GeV) \tabularnewline\hline
500.00 & -482.09 & 749.68 & 596.69 \tabularnewline\hline
1000.00 & -550.00 & 939.89 & 598.67 \tabularnewline\hline
1500.00 & -650.00 & 1195.17 & 598.65 \tabularnewline\hline
2000.00 & -800.00 & 1484.31 & 595.54 \tabularnewline\hline
2500.00 & -1050.00 & 1789.46 & 600.00 \tabularnewline\hline
3000.00 & -1350.00 & 2105.50 & 601.05 \tabularnewline\hline
3500.00 & -1700.00 & 2427.71 & 601.96 \tabularnewline\hline
4000.00 & -2080.00 & 2754.36 & 599.09 \tabularnewline\hline
4500.00 & -2500.00 & 3083.11 & 600.12 \tabularnewline\hline
5000.00 & -2950.00 & 3413.30 & 605.43 \tabularnewline\hline
\end{tabular} 
\caption{\label{tab_a0_m0} 
 Display of HB/FC1 for $m_{1/2}=400\GeV$ and $\tan\beta=15$.  This is an example of HB/FC1 with $A_0<0$ solution with $\mu=\left(600\pm6\right)\GeV$.
The values of $\mu$ have been calculated with both {\sc SuSpect}~\cite{SuSpect} and  {\sc SoftSUSY}~\cite{softsusy}. 
} 

 \end{center} 
  \end{table}

 Now there is also another possibility of achieving a Focal~Curve   
 which can be illustrated by writing Eq.~\eqref{1.6} in the form 
   \beqn
  \mu^2 + \frac{1}{2} M_Z^2 =  m_0^2 C_1 +  A_0^2 \overline{C}_2 + \overline{m}_{1/2}^{2} C_3  + \Delta \mu^2_{\rm loop}~.  \\
\label{4.6}
\overline{m}_{1/2} \equiv m_{1/2} + \frac{C_4}{2C_3} A_0, ~~~\overline{C}_2 \equiv C_2 -\frac{C^2_4}{4C_3}~.
\label{4.7}
\eeqn
As before, we can write this equation in the form
\beqn
 \left(\sqrt {C_3} \overline{m}_{1/2}\right)^2   - \left(\sqrt{|C_1|} m_0\right)^2 = \pm |\mu_2|^2~~~~{\rm HB/FC2}~,~~
\label{4.8}
\eeqn
where 
$\pm |\mu_2|^2   \equiv \mu^2 + \frac{1}{2} M_Z^2 -  A_0^2 \overline{C}_2 - \Delta \mu^2_{\rm loop}$~.
Thus again one has two  branches depending on the sign.
Here one keeps $A_0$ fixed while $m_0$ and $m_{1/2}$ get  large and $\mu$ is 
relatively small~(this is Focal~Curve HB/FC2 as defined in Sec.~\ref{s2}).  
  For the case when $|\mu_2|$ is small relative to $m_0$ and $\overline{m}_{1/2}$ one finds
the following relationship asymptotically
 \beqn
\frac{\overline{ m}_{1/2}}{m_0}   \longrightarrow ~ \frac{m_{1/2}}{m_0} \longrightarrow  \sqrt{\frac{|C_1|}{C_3}}  \longrightarrow \approx  0.4~.
  \label{4.5a}
  \eeqn
  where the last entry in Eq.~\eqref{4.5a} is obtained by using Eq.~\eqref{1.61} and Eq.~\eqref{1.61b}
  as shown in Fig.~\ref{figc12c13}.  
    An  illustration of this case is given in the right panel of Fig.~\ref{a0_m0_nl} where  $\mhf$ 
gets very large.  For these curves we see that we can still have models with $\mu$ small ($\mu\lesssim 450\GeV$) and $\mhf$ large~($\mhf\gtrsim 1500\GeV$), which leads to the gluino mass being on the order of a few TeV or larger.

  To show that there exists a larger set of Focal~Curves 
  than the cases we have discussed above
  we exhibit a whole set of parametric Focal~Curves which we label as HB/FC$_{\alpha}$. To do this we define $\left(1-\alpha\right)\mhf=\alpha|A_0|$,
  where $0< \alpha <1$.     
This allows us to rewrite Eq.~\eqref{1.6} as
\beqn
\pm |\mu_{\alpha}|^2=-\left(\sqrt{|C_1|}\right)^2  m_0^2   + C_{\alpha} A_0^2~.
\label{433}
\eeqn
where
$\pm |\mu_{\alpha}|^2=\mu^2+\frac12 M_Z^2-\Delta\mu_{\rm loop}^2$.
Further, 
\beqn
C_{\alpha} = C_2+\frac{\alpha^2}{\left(1-\alpha\right)^2} C_3+\frac{\alpha }{1-\alpha}C_4  
~{\rm sgn}\left(A_0\right)~,
\label{433a}
\eeqn 
Eq.~\eqref{433} shows that there exists parametric Focal~Curves, parameterized by $\alpha$, where 
one can get the same value of $\mu$ which can be taken to be small, while $\alpha$ can take on 
values in the range $(0,1)$. This phenomenon illustrated in the right panel of
Fig.~\ref{alph_fig} displays several Focal~Curves for constant $\mu$. 
One finds that as $\alpha$ decreases
 the asymptotic form of the curves in the $A_0-m_0$ plane become 
more steep. This result is in agreement with the  theoretical prediction 
at one loop for the  asymptotic ratio $A_0/m_0$  which is 
  \beqn
A_0/m_0\rightarrow  \pm \sqrt{\left|C_1\right|/C_{\alpha}}~.
  \eeqn
{\scriptsize
 \begin{table}[t!]
\begin{center}
\scriptsize
{\bf Focal regions  in mSUGRA Hyperbolic Branch} \\~\\
\begin{tabular}{|c|c|c|c|}
\hline
 Focal Region & Symbol  & varying parameters  & fixed parameters \\
\hline
Focal~Point & HB/FP &  $m_0$ & $m_{1/2}, A_0$ \\
Focal~Curve& HB/FC1 & $m_0,A_0$  & $m_{1/2}$ \\
Focal~Curve& HB/FC2 & $m_0,m_{1/2}$  & $A_0$ \\
Focal~Curve& HB/FC$_{\alpha}$ & $m_0, A_0~~{\rm or}~~\mhf$ & $m_{1/2}= \frac{\alpha}{1-\alpha} |A_0|$  \\
Focal Surface &HB/FS &  $m_0, m_{1/2}, A_0$ &     \\  
  \hline
\end{tabular}
\caption{A summary of the classification of focal regions in mSUGRA.
The focal regions are those where $\mu$ remains constant while 
one or more soft parameters may get large. $\tan\beta$ is assume fixed 
in each of the cases discussed and $\alpha$ has the range $0< \alpha <1$.
\label{tab_msugra}}
\end{center}
\end{table} 
}

\begin{figure}[t!]
\begin{center}
{\scriptsize \bf Focal Surface HB/FS}\\ 
\includegraphics[scale=0.12]{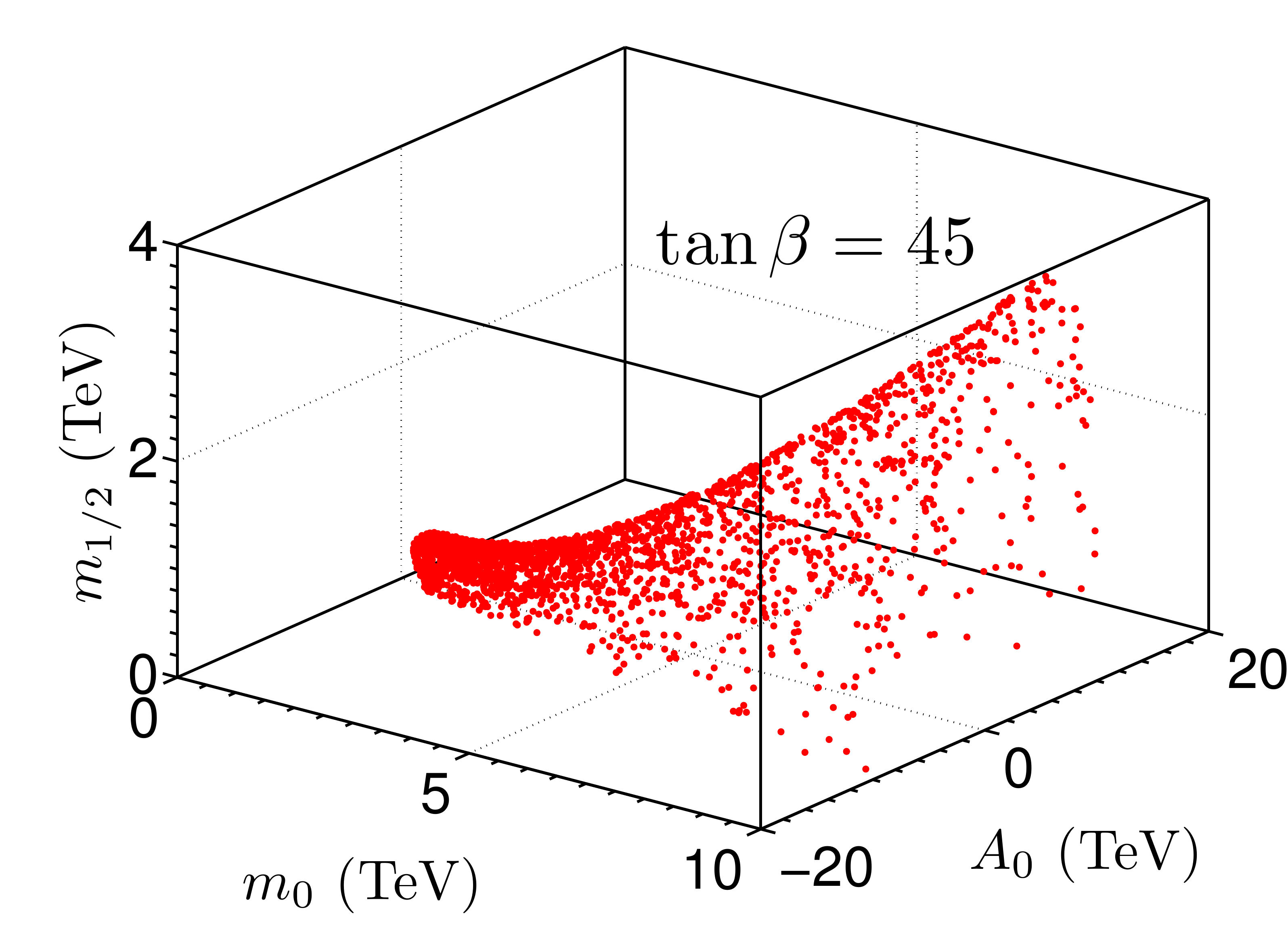}
\includegraphics[scale=0.18]{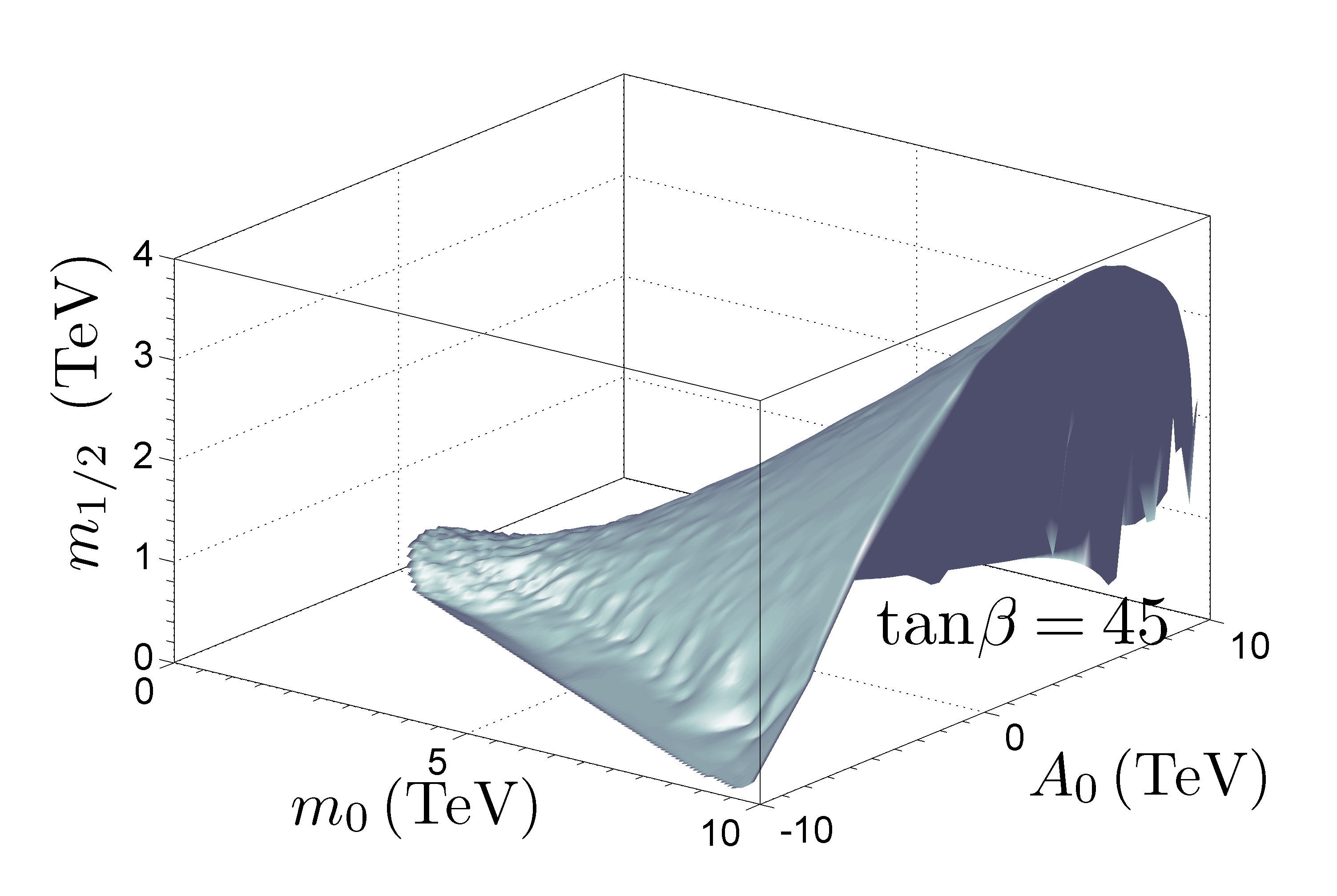}
\caption{\label{null_surf_fig} 
Exhibition of a Focal Surface 
 when $\tan\beta =45$ and $\mu=\left(0.465\pm0.035\right)\TeV$ while 
$m_0, m_{1/2}, A_0$ can  all get large. The left panel shows a scatter plot of model points lying on a Focal Surface.
The right panel  shows the same Focal Surface using an interpolation of the points presented in the left panel.
  } 
\end{center}
\end{figure}

{\it Focal Surfaces HB/FS:}
We consider next  the radiative breaking of the electroweak symmetry 
 where  all the three parameters
$m_0$,  $\mhf$, or $A_0$  can get large while $\mu$ remains small.
This solution is again valid in the region of the parameter space
where $C_1$ turns negative at the value of renormalization group point which minimizes
the loop correction. This is the Focal Surface HB/FS as defined in Sec.~\ref{s2} and we can express
it in the following two forms
\beqn
&\pm& |\mu_s|^2=-\left(\sqrt{|C_1|} m_0\right)^2 + \left(\overline{A}_0\sqrt {C_2} \right)^2 + \left(\sqrt{\overline{C}_3} m_{1/2}\right)^2 
\nonumber
\eeqn
where $\pm |\mu_s|^2 =\mu^2 + \frac{1}{2}M_Z^2-\Delta\mu_{\rm loop}^2$. 
A summary of focal~regions is given in Table~\ref{tab_msugra}.
An exhibition of a Focal Surface for the case $\mu=(0.465 \pm 0.035)\TeV$ is given 
 in Fig.~\ref{null_surf_fig}. We note that on the Focal Surface shown in Fig.~\ref{null_surf_fig}
  $m_0$,  $\mhf$, or $A_0$ can all be seen to get large in certain regions while $\mu$
  remains relatively constant.  
   We note in passing that another way to generate a Focal Surface is to consider a Focal Curve 
  HB/FC$_{\alpha}$ and let $\alpha$ vary over its allowed range $0\leq \alpha <1$.
Thus a Focal Surface can be viewed as a
 collection  of Focal Curves as in the right panel of Fig.~\ref{alph_fig}.  
  
\section{LHC and Dark Matter Implications\label{s7}}
{\it Constraints of LHC-7 data on Focal Regions:} 
We now investigate the implications of the  recent  LHC data~\cite{cmsREACH,AtlasSUSY,atlas0lep,atlas165pb,atlas1fb} on the focal regions constituted of the Focal Point, Focal Curves and Focal Surfaces.
To this end we first generate 
 mSUGRA parameter points using a uniformly distributed random scan over the soft parameters with $m_0 < 4\TeV$, $m_{1/2} < 2\TeV$, $A_0/m_0 \in (-10,10)$, and $\tan\beta \in(1,60)$. 
After the constraint of  REWSB roughly 22~million mSUGRA
 parameter points are collected.
These are then subject to experimental constraints which  include the LEP and Tevatron~\cite{pdgrev} limits on
the Higgs mass and on the sparticle masses as discussed in~\cite{Akula:2011dd,Akula:2011ke}
and
${\mathcal{B}r}\left(B_{s}\to \mu^{+}\mu^-\right)\leq 1.1\times10^{-8}$~\cite{cmslhcbbsmumu}.
These  constraints will be referred to as the {\it general constraints}.
In imposing these constraints we use 
  {\sc MicrOMEGAs}~\cite{micro} for the computation of the  relic density  
and {\sc SuSpect} for the computation of the sparticle mass spectrum and $\mu$ at the scale at which electroweak symmetry breaks, $Q_{\rm EWSB}$. A more statistically rigorous procedure for the implementation of the constraints would be  to use $\chi^2$ or maximum likelihoods, but for the purpose of this analysis it is unnecessary.

\begin{figure}[t!]
\begin{center}
\begin{tabular}{cc}
 {\scriptsize\bf EB} & {\scriptsize\bf HB/FP + HB/FC}\\
\includegraphics[scale=0.1]{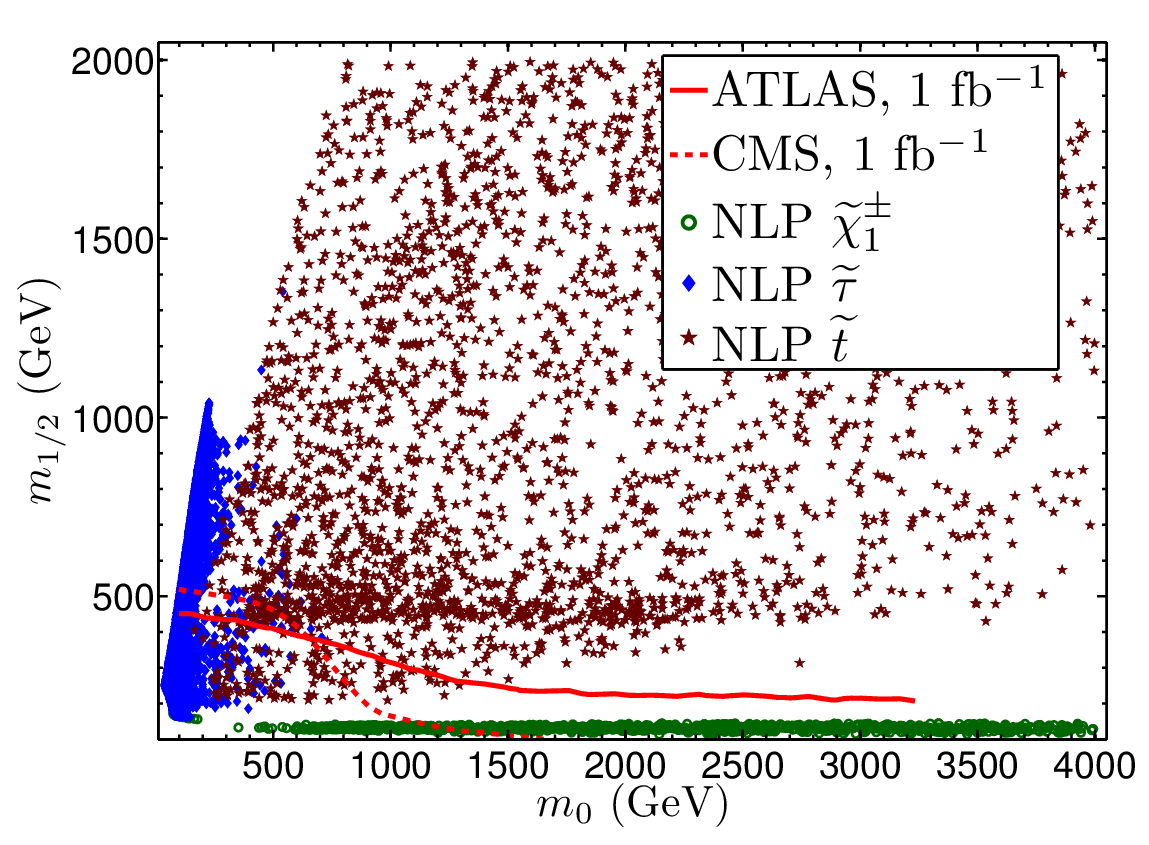}
 & \includegraphics[scale=0.1]{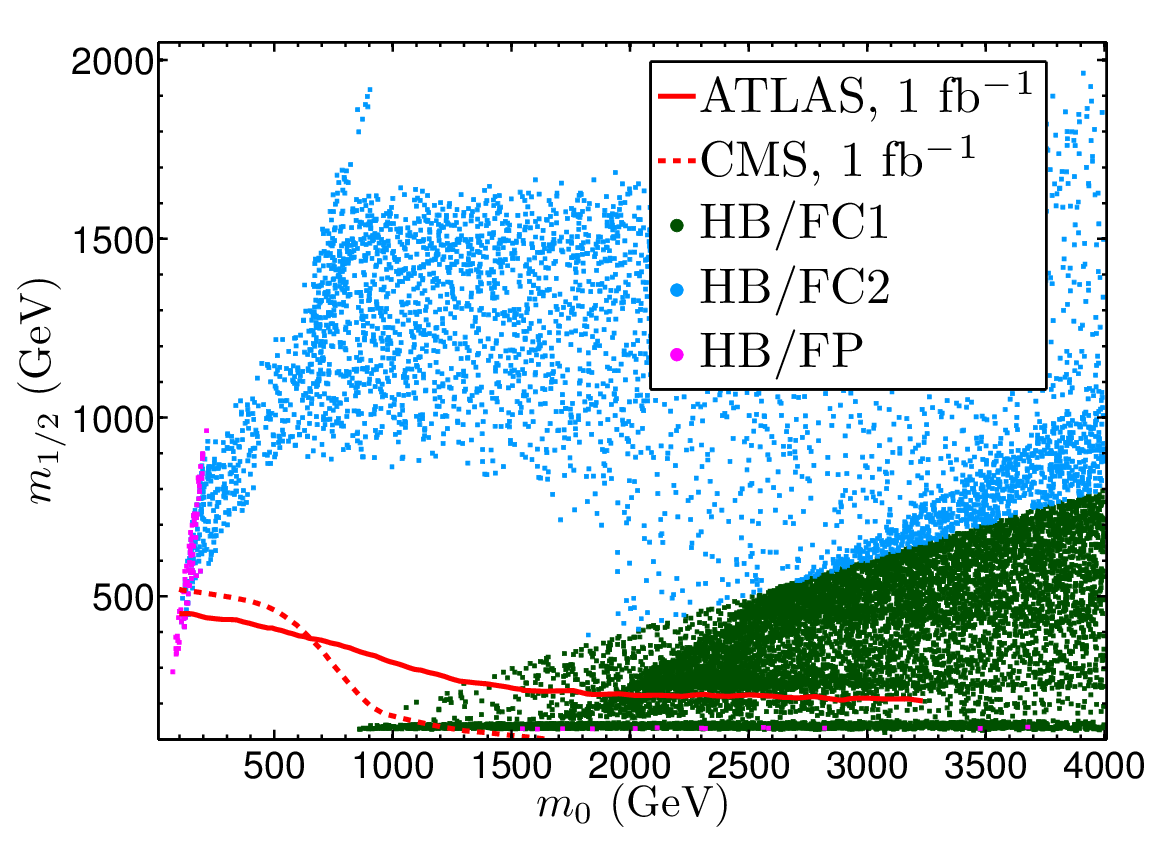}\\
{\scriptsize\bf HB/FP + HB/FC} & {\scriptsize\bf HB/FP}\\
\includegraphics[scale=0.1]{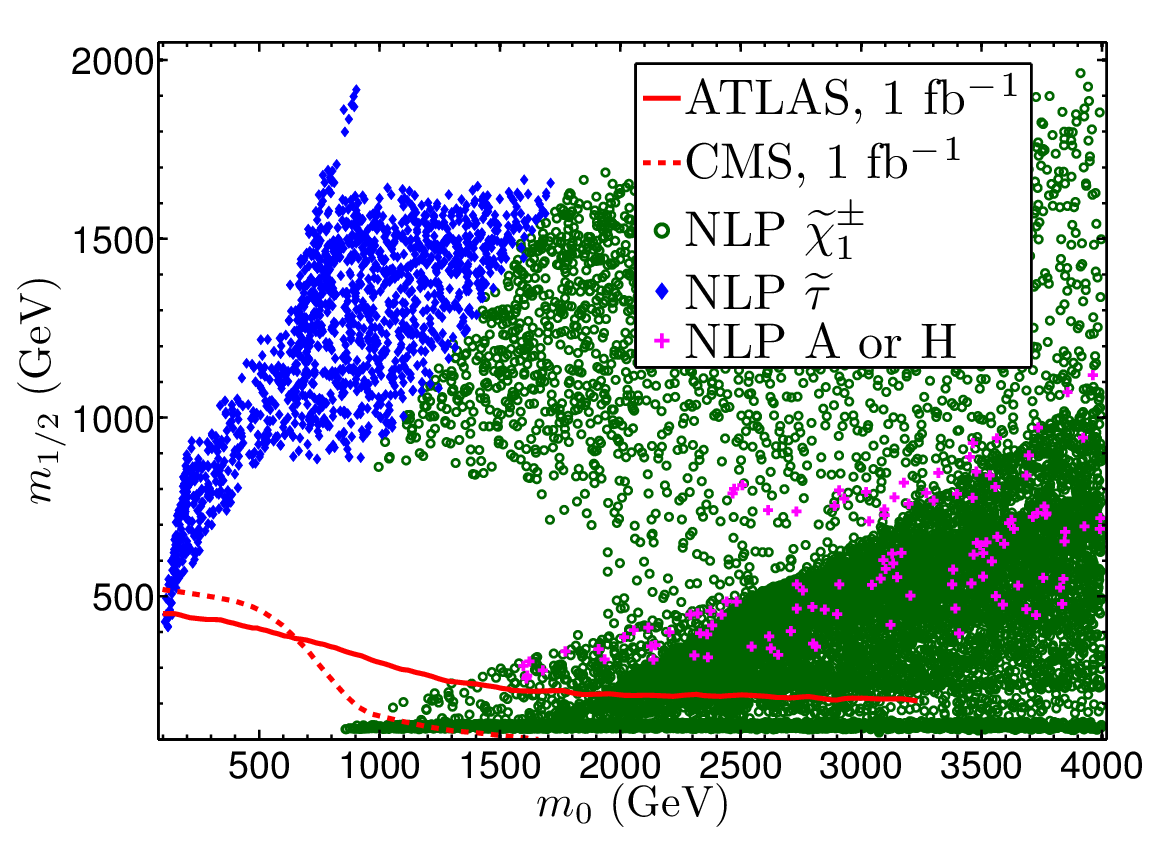}
&\includegraphics[scale=0.1]{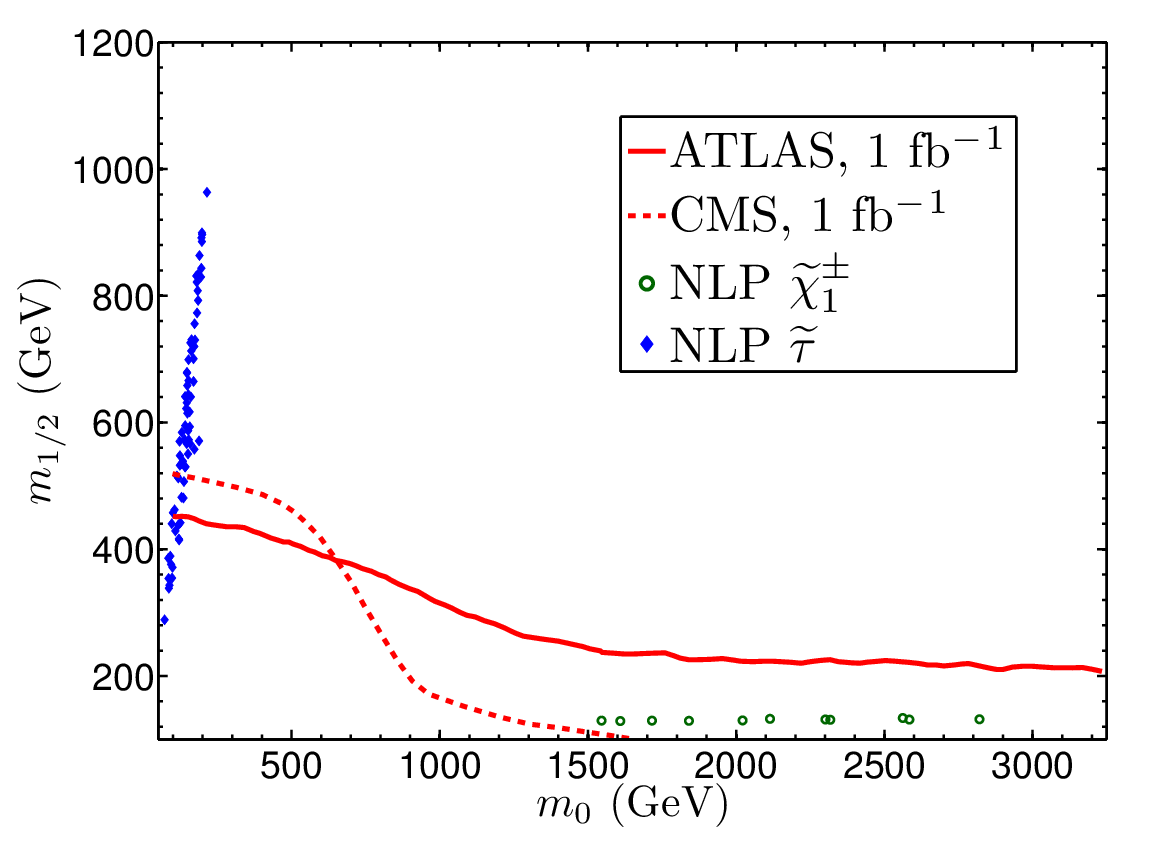}\\
\multicolumn{2}{c}{\bf \scriptsize HB/FS}\\
\includegraphics[scale=0.1]{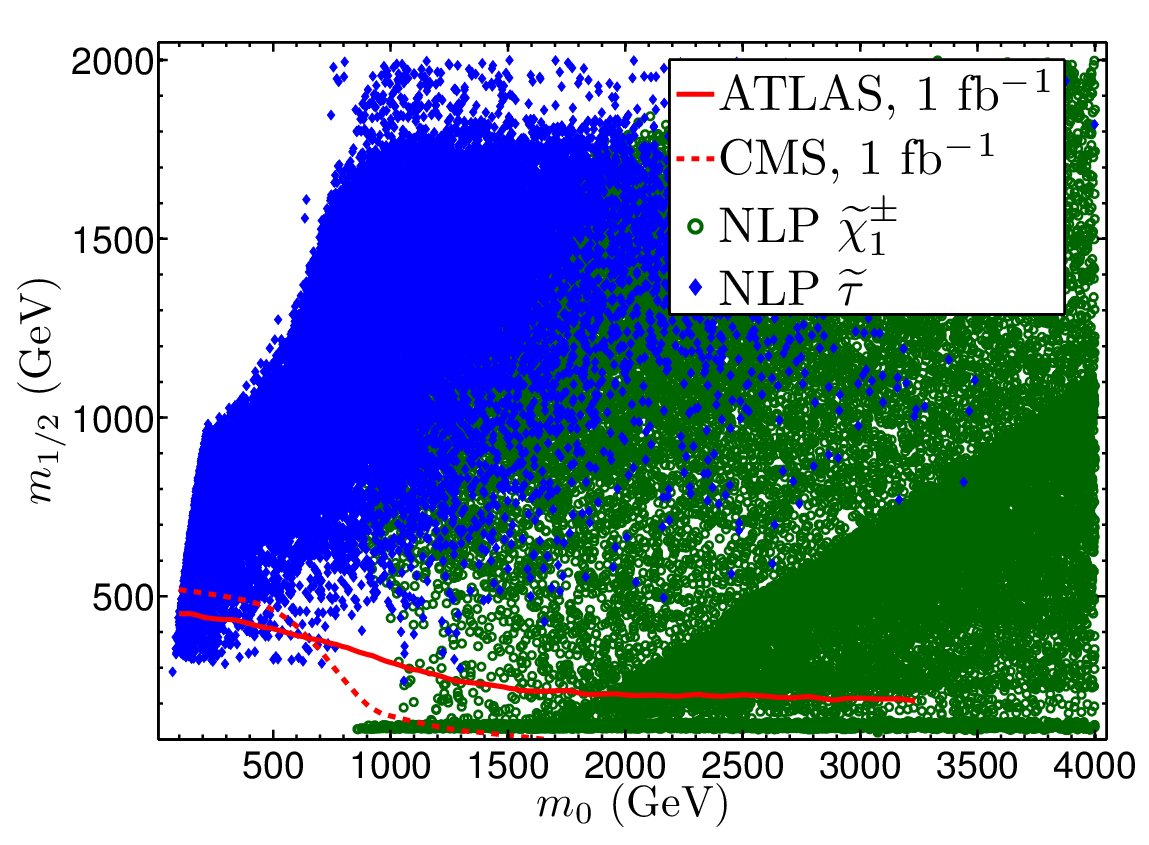}
&\includegraphics[scale=0.1]{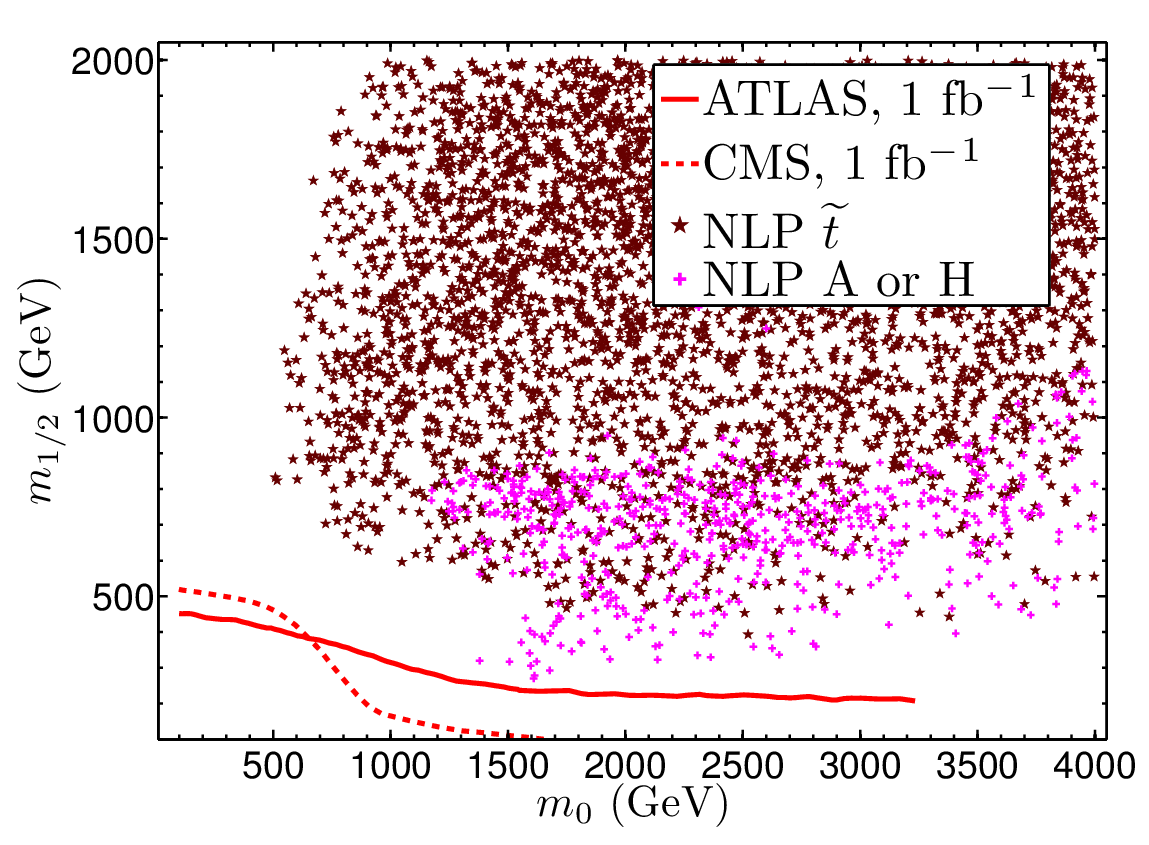}
\end{tabular}
\caption{\label{m0mhf_msugra2}
Top Left: The mSUGRA parameter points passing the
general constraints in the $m_0$--$m_{1/2}$ plane that are a part of
the Ellipsoidal Branch, labeled by the NLP where in the definition of EB we have excluded the HB/FP region.
Top Right: The mSUGRA parameter points in the $m_0-m_{1/2}$
plane passing the general constraints that are a part of HB/FC1,
HB/FC2, or HB/FP, labeled as such.
Middle Left : An exhibition of the mSUGRA parameter
points passing general constraints that also lie on HB/FC1 or HB/FC2, labeled by the NLP.
Middle Right: The mSUGRA parameter points passing
the general constraints that arise from the Focal Point (HB/FP) region.
Bottom Left: 
A display of the mSUGRA parameter points containing the $\cha$ and
the $\sta$ NLPs  
passing the general constraints 
and including the parameters in HB/FS, i.e., the entire
Hyperbolic Branch except for HB/FP.  
Bottom Right: Same as bottom left except the NLPs displayed are $\ta$,
$A, H$.
}
\end{center}
\end{figure}

CMS and ATLAS have reported results for supersymmetry searches~\cite{cmsREACH,AtlasSUSY,atlas0lep,atlas165pb,atlas1fb} based on about $1~\fb$ of data. The implications of these results (as well as dark matter results)
have been considered for the  parameter space of SUSY models 
in a number of works~\cite{Farina:2011bh,Akula:2011zq,LHC7,Akula2,Akula:2011dd,Grellscheid:2011ij} and some discussion on the
collider implications on naturalness can be found in~\cite{Ellwanger:2011mu,Cassel:2010px, Papucci:2011wy,Gogoladze:2009bd}.
Here we use the constraint arising from the recent  ATLAS $1~\fb$ search~\cite{atlas1fb} and the CMS $1~\fb$ search~\cite{cmsREACH} to explore their implications on the  focal region. 
The implications of the LHC data for the Ellipsoidal Branch and for the Hyperbolic Branch are exhibited in
Fig.~\ref{m0mhf_msugra2}.    The top left panel gives the parameter space in  EB and here one finds that
most of the model points being constrained by LHC-7 lie in the low $m_0$ region. The top right panel
gives the corresponding analysis for HB/FP and HB/FC. In the analysis here we have assumed that
$m_{1/2}/m_0\leq 0.1$ for HB/FC1 and $A_0/m_0 \leq 0.1$ for HB/FC2.
The middle  left panel exhibits the same set of parameter points
on HB/FP and HB/FC as the top left panel except that the regions are now labeled according to the sparticle landscape 
picture~\cite{landscape}
by the next to lightest particle beyond the Standard Model (NLP) in the mass hierarchy (note that this includes all of the sparticles and Higgs sector particles, but omits the Standard Model-like $h^0$). Here one finds that most of the region being constrained
by the LHC-7 data is the high $m_0$ region.  
The middle right panel exhibits the Focal Point region, HB/FP.
Here one finds that the Focal Point region HB/FP is highly depleted
and is further constrained by the LHC-7 data. 
The bottom panels of Fig.~\ref{m0mhf_msugra2} show the parameter points on HB/FS which is the entire HB region except the HB/FP region.  The left panel displays the parameter points 
where the NLP is either a $\cha$ or $\sta$, and the right hand panel shows the 
parameter points where the NLP is $\tilde t, A$ or $H$.  
 Thus the analysis of Fig.~\ref{m0mhf_msugra2} shows that the 
HB/FP is almost empty and  most of the parameter space remaining on HB lies in the region of Focal Curves or Focal Surfaces, i.e., it lies on HB/FC and HB/FS.

\begin{figure}[t!]
\begin{center}
{\bf \scriptsize  $\cha\nb$ Production at LHC-7 on HB/FC1}\\
\includegraphics[scale=0.18]{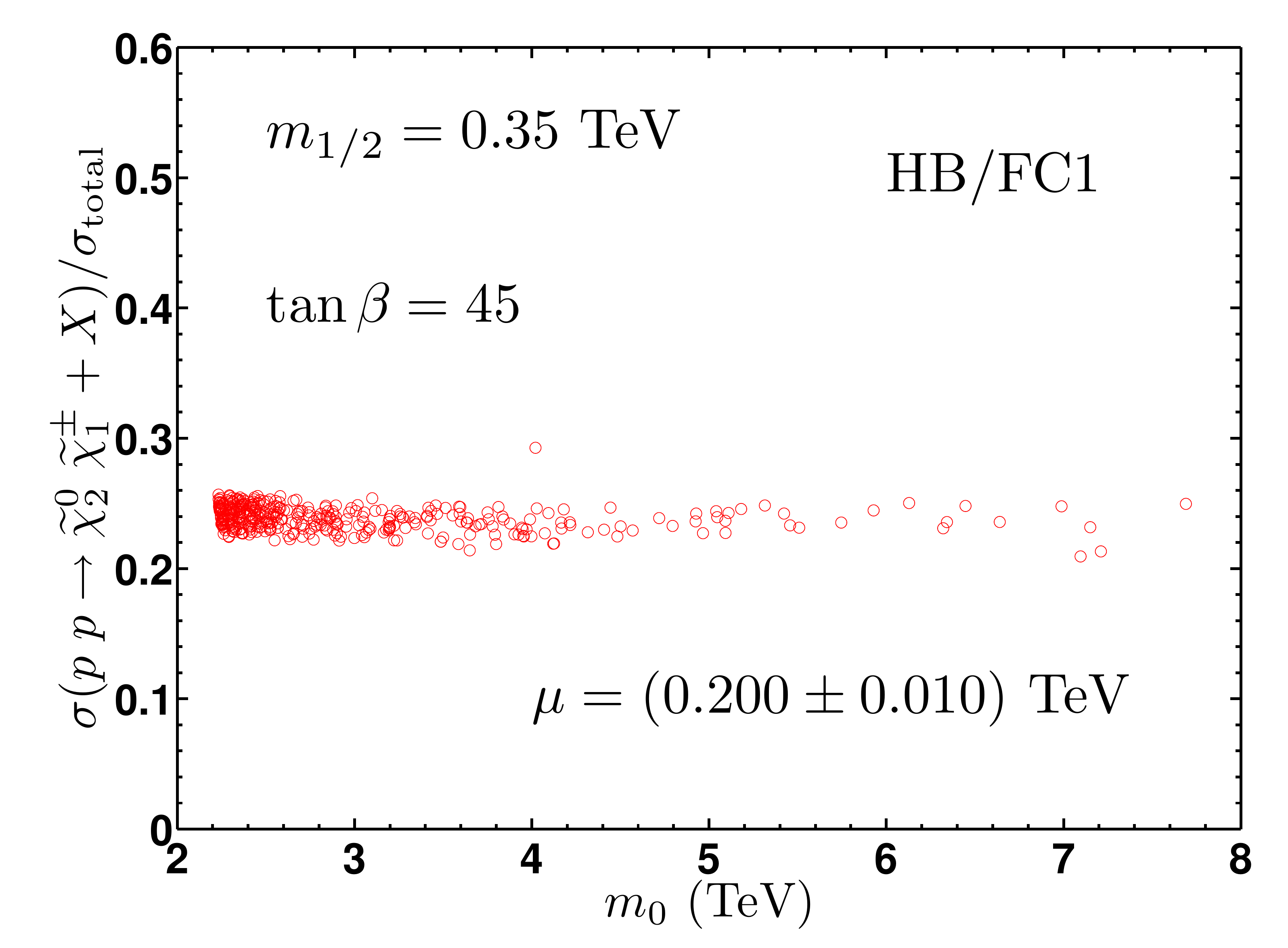}
\caption{\label{nl_lhc}
 Fraction of the total cross-section that is made up by $\cha\nb$ production as a function of $m_0$
at  $\sqrt{s}=7\TeV$. The analysis shows that the production cross-section is rather 
insensitive to $m_0$ which implies the signatures from HB/FC1  such as the 
trileptonic signal could be visible even in the asymptotic region when $m_0$ and $A_0$ are very large.  
}
\end{center}
\end{figure}

{\it LHC signals on HB/FC1: }  We discuss now an important phenomenon related to 
HB/FC1, 
which arises from the constraint that $\mhf$ and $\mu$ are fixed even though $A_0$ and $m_0$ 
get large.  This can lead to  observable leptonic signatures, specifically the trileptonic signature~\cite{bht,trilep}, even when $m_0$ lies in 
the several TeV region (For a recent work on the trileptonic signal see~\cite{Bornhauser:2011ab}). The reason for this is rather obvious, in that the 
 chargino and the neutralino masses are held relatively constant along the Focal Curve HB/FC1. 
 Thus the production cross-section for the charginos and neutralinos will be essentially independent of $m_0$. 
 We are specifically interested in the production cross-section of the light chargino $\cha$ and the second lightest neutralino
 $\nb$, i.e.,   $\sigma_{\cha\nb}$ which can lead to a trileptonic  signal from the decay of $\cha, \nb$ so that
 $\cha\to l^{\pm} + \nu_l+ \na$ and $\nb\to l^+l^- \na$ (important contributions can also arise from 
 the production of $\cha\noi$ (i=3,4) depending on the part of the parameter space one is in).  
 The chargino and neutralino  final state can arise at tree level from two main processes in $pp$ collisions.
 Thus, for example, $\chpa \noi$ can arise from the s-channel fusion diagram $u+\bar d\to W^{+*}\to \chpa + \noi$ and from the t-channel exchange diagram of a $\tilde d_L$ squark. The latter diagram is suppressed
 when $m_0$ is large so that the main production cross-section proceeds via the s-channel off -shell
 $W^{\pm}$ production~\cite{trilep}. Thus the $\cha \noi$ production cross-section is expected to be independent
 of $m_0$ for large $m_0$.   
  The constancy of $\sigma_{\cha\nb}/\sigma_{\rm total}$ is exhibited in  Fig.~\ref{nl_lhc} for HB/FC1 defined by
  $m_{1/2} =0.35\TeV$,  $\tan\beta=45$ and $\mu=(0.20 \pm 0.01)\TeV$. 
 The branching ratio into trileptons is also 
  computed.
     In the analysis we use  {\sc SUSY-HIT}~\cite{susyhit}  for the computation of decays, 
   {\sc PYTHIA}~\cite{pythia} for event generation,  and {\sc PGS}~\cite{pgs} for detector simulation.
For the case of models exhibited in Fig.~\ref{nl_lhc}  the $\cha\nb$ production
cross-section is $\left(164.3\pm 9.97\right)~{\rm fb}$ and the $\cha\nc$ production 
cross-section is $\left(112.1\pm 8.53\right)~{\rm fb}$, which leads to roughly $50$~raw trilepton events at $10~\fb$ 
 where we have included $\tau$s in the definition of leptons.  The number of events will be reduced when off-line cuts are imposed and a more detailed analysis would require further knowledge of the cuts used in the experimental multileptonic search at that luminosity. Of course a much larger number of events is expected at 
 higher $\sqrt s=10\TeV$, or $\sqrt s=14\TeV$ at the same luminosity. Similarly, the $\cha\nb$ and $\cha\nc$ production states can decay hadronically.  For the hadronic analysis we use the cuts as outlined in~Ref.~1~of~\cite{atlas1fb} by ATLAS and find that our effective cross-sections are $\left(5.2\pm 0.15\right)~{\rm fb}$, $\left(0.7\pm 0.16\right)~{\rm fb}$, $\left(1.6\pm 0.33\right)~{\rm fb}$, $\left(0.6\pm 0.18\right)~{\rm fb}$ and $\left(0.5\pm 0.15\right)~{\rm fb}$ which can be compared to the reported $95\%$~C.L. upper bounds at $1.04~\fb$ of $22~{\rm fb}$, $25~{\rm fb}$, $429~{\rm fb}$, $27~{\rm fb}$ and $17~{\rm fb}$, respectively.  
Typically these points produce hard jet signatures, but with low jet multiplicity.  
Thus the hadronic signals on HB/FC1 may become visible if a luminosity in excess of $20~\fb$ can be achieved at LHC-7.  Another possible channel for discovery would be a combination of jets and leptons, but such an analysis is outside the scope of the current work.

\begin{figure}[t!]
\begin{center}
\begin{tabular}{cc}
\multicolumn{2}{c}{\bf \scriptsize Full mSUGRA parameter space}\\
\includegraphics[scale=0.1]{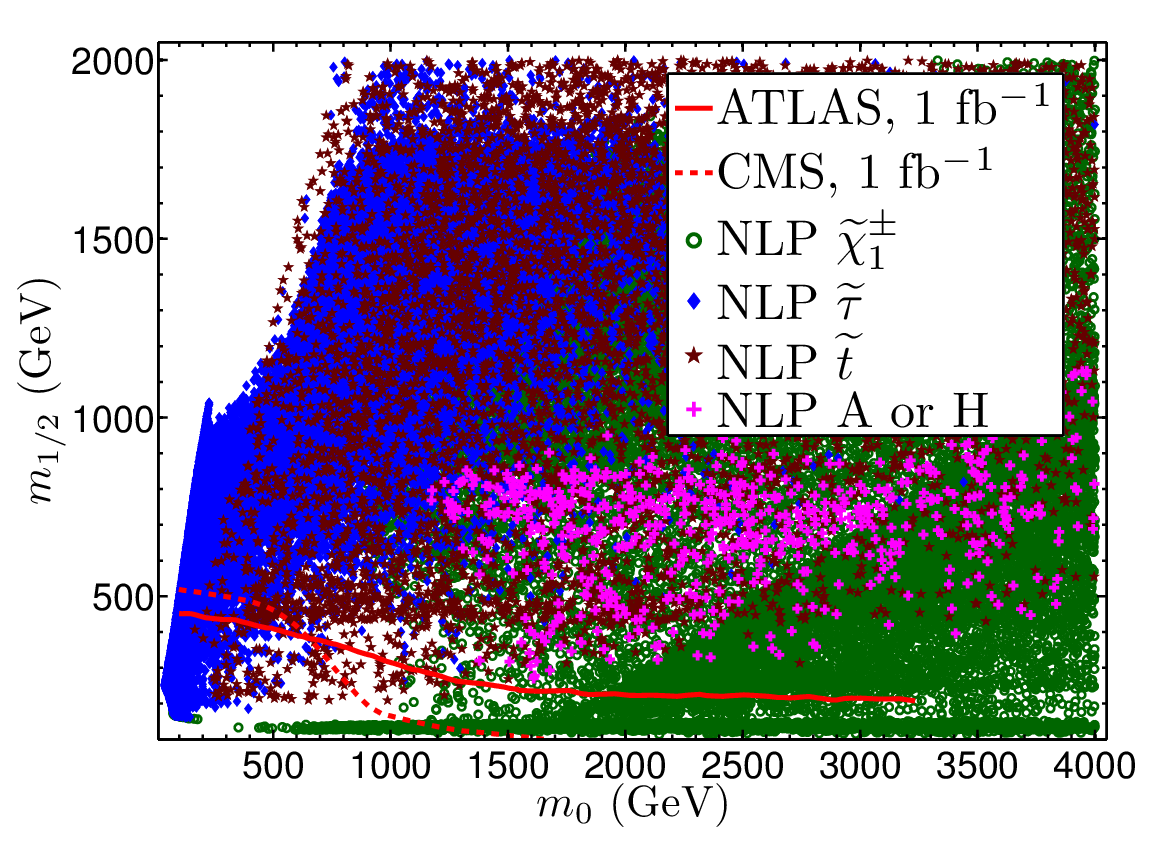}
&  \includegraphics[scale=0.1]{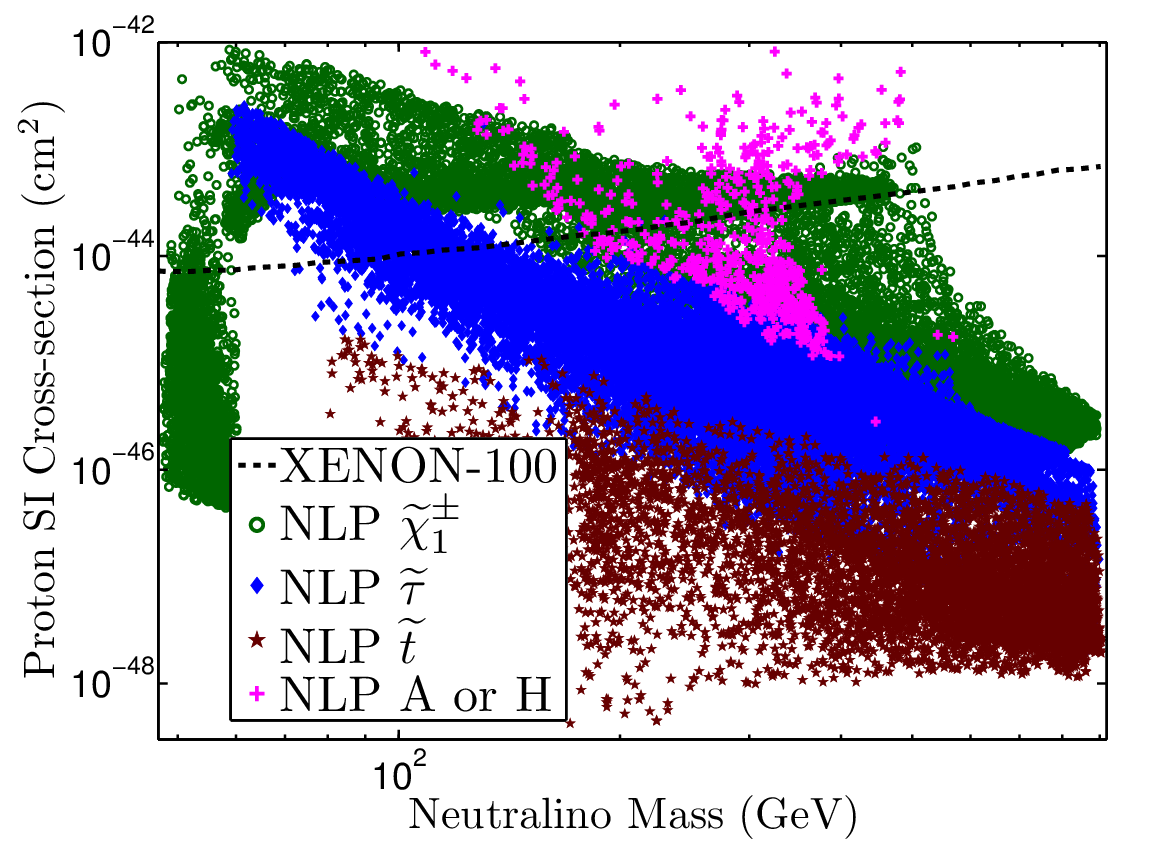}\\
  {\bf \scriptsize Dark Matter on EB} & {\bf \scriptsize Dark Matter on HB}  \\
\includegraphics[scale=0.1]{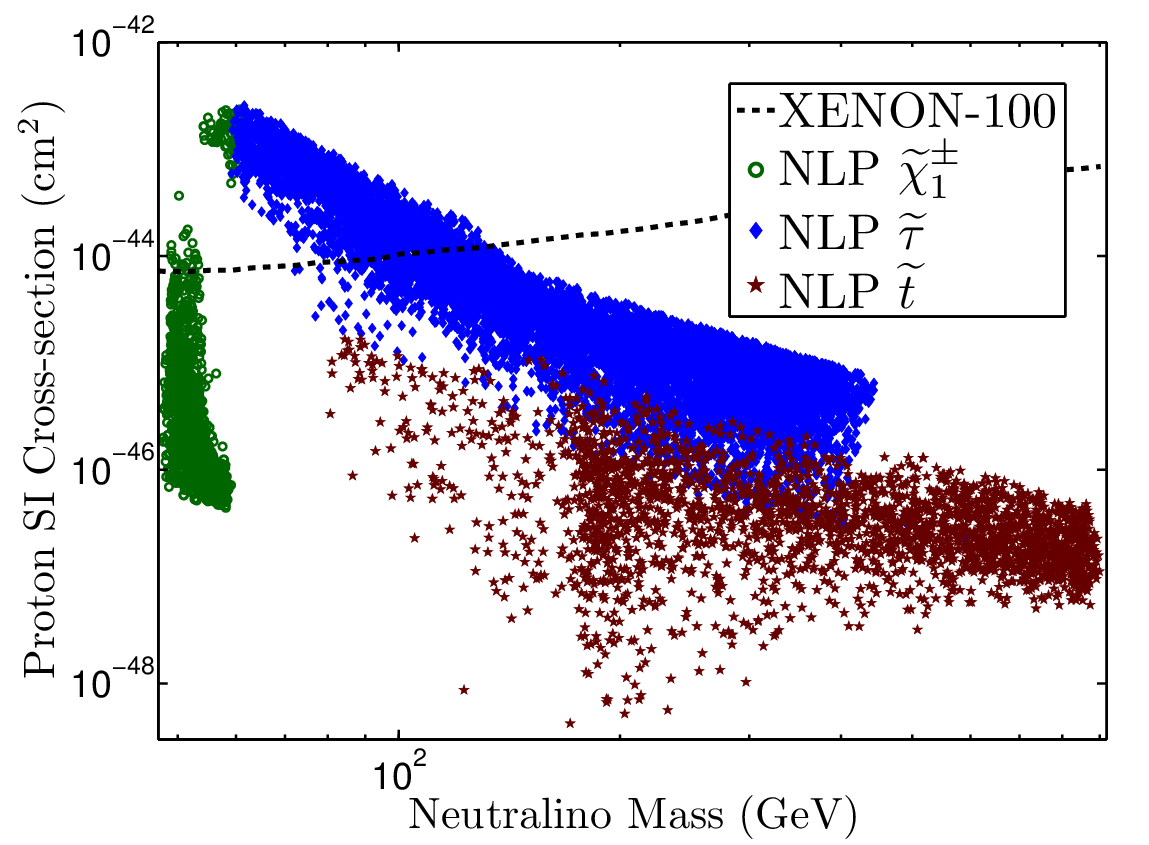}
& \includegraphics[scale=0.1]{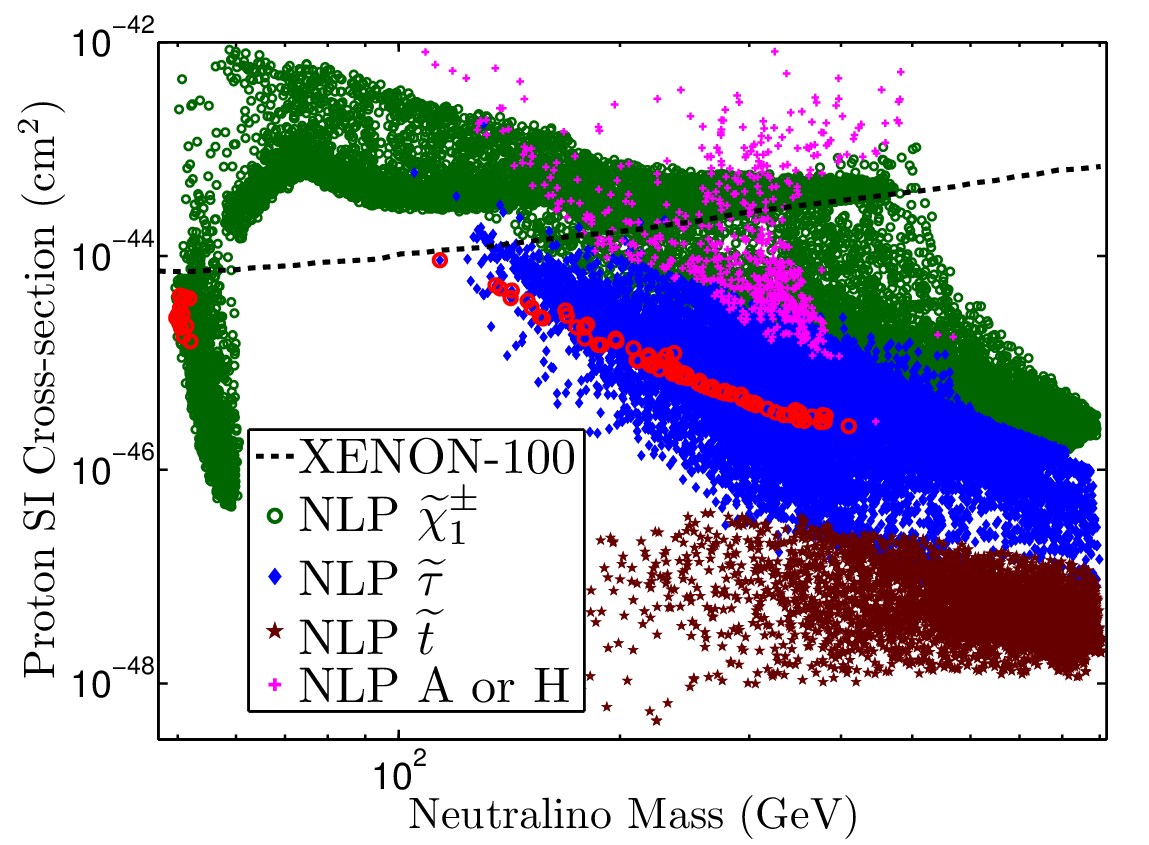}
\end{tabular}
\caption{\label{fig_dm}
Top left: 
A display for the mSUGRA model points in the $m_0-m_{1/2}$ plane
that pass the general constraints as discussed in the text.
Top right: A display of the  spin-independent  neutralino-proton cross-section $\sigma_{\na,p}^\mathrm{SI}$
 for the parameter points in the top left panel. 
 Bottom Left: A display of the  spin-independent  neutralino-proton cross-section, $\sigma_{\na p}^\mathrm{SI}$,
 for the EB region. Bottom Right:  Same as the bottom left except that the analysis is for 
 HB which contains the Focal Point as well as Focal Curves and Focal Surfaces.
}
\end{center}
\end{figure}

{\it Dark Matter in the EB and the Focal Domains:}
It is interesting to investigate the prediction for dark matter searches in EB vs HB
 domains. 
We begin by considering first the full parameter space of mSUGRA which, after general constraints,
is exhibited in the top left panel of Fig.~\ref{fig_dm}  where the  LHC-7 constraint with $1~\fb$ of data is also exhibited. The spin-independent cross-section vs the neutralino mass corresponding to the parameter space
in the top left panel is exhibited in the top right panel where we have also exhibited the 
experimental exclusion from XENON-100 experiment. 
Next, in the bottom left panel of Fig.~\ref{fig_dm}, we 
exhibit the spin-independent neutralino-proton cross-section vs the neutralino mass 
for EB while the bottom right panel exhibits the same for the full HB domain consisting of  HB/FP, HB/FC and HB/FS. 
The  HB/FP region indicated  by the red area 
is rather small while most of the remaining parameter
space is constituted of HB/FC and HB/FS.


\section{Conclusion\label{conclusion}\label{s8}}
It is shown that the Hyperbolic Branch of radiative electroweak symmetry breaking  
consists of several regions of the parameter space where $\mu$ is small. These
regions consist of the Focal~Points, Focal~Curves and Focal~Surfaces. The Focal~Point~(HB/FP) region is 
where  $m_0$ can get large with fixed $m_{1/2}$ and $A_0$ while $\mu$ remains small.
A small $\mu$ can also be achieved on Focal~Curves and on Focal~Surfaces. 
There are two possible Focal~Curves: HB/FC1 and HB/FC2 such that on HB/FC1, 
  $m_0$ and $A_0$ both may get  large, while $m_{1/2}$ and $\mu$  remain fixed,
 while on HB/FC2, $m_0$ and $m_{1/2}$ may get large while $A_0$ remains fixed. 
 These two general categories can be unified by the parameter $\alpha$ defining the 
 Focal Curve mode $\mathrm{HB/FC}_\alpha$.
 An explicit illustration of these regions is given for mSUGRA where it is shown 
 that the HB/FP region is significantly depleted 
 when all the experimental constrains along with the current constraints from the LHC-7 data are applied. Thus the remaining
 parameter points in this region lie on Focal~Curves (or more generally, on Focal~Surfaces).
 Thus if $m_0$ is indeed large while the gaugino masses are light, the LHC would turn into a gaugino factory.  Some of the SUSY signals for this case were also discussed. 
 
\begin{acknowledgements}
This research is  supported in part by grants    PHY-0757959 and PHY-0969739,  
and by  TeraGrid  grant TG-PHY110015. GP acknowledges TASI 2011 where a portion of this work was completed.
\end{acknowledgements}
%
%
%
%

%
%
%
%
\end{document}